\documentclass[12pt,preprint]{aastex}

\usepackage{pdflscape}

\slugcomment{To appear in ApJ}

\shorttitle{Fullerene PNe}
\shortauthors{Garc\'{\i}a-Hern\'andez et al.}

\begin{document}


\title{Infrared Study of Fullerene Planetary Nebulae}


\author{D. A. Garc\'{\i}a-Hern\'andez\altaffilmark{1,2}, E.
Villaver\altaffilmark{3}, P. Garc\'{\i}a-Lario\altaffilmark{4}, J. A.
Acosta-Pulido\altaffilmark{1,2}, A. Manchado\altaffilmark{1,2,5}, L.
Stanghellini\altaffilmark{6}, R. A. Shaw\altaffilmark{6} and F.
Cataldo\altaffilmark{7,8}}


\altaffiltext{1}{Instituto de Astrof\'{\i}sica de Canarias, C/ Via L\'actea
s/n, E-38205 La Laguna, Spain; agarcia@iac.es, amt@iac.es}
\altaffiltext{2}{Departamento de Astrof\'{\i}sica, Universidad de La Laguna
(ULL), E-38206 La Laguna, Spain}
\altaffiltext{3}{Departamento de F\'{\i}sica Te\'orica, Universidad
Aut\'onoma de Madrid, E-28049 Madrid, Spain; eva.villaver@uam.es}
\altaffiltext{4}{Herschel Science Centre. European Space Astronomy Centre,
Research and Scientific Support Department of ESA. Villafranca del Castillo,
P.O. Box 78. E-28080 Madrid. Spain; Pedro.Garcia-Lario@sciops.esa.int }
\altaffiltext{5}{Consejo Superior de Investigaciones Cient\'{\i}ficas, Spain}
\altaffiltext{6}{National Optical Astronomy Observatory, 950 North Cherry
Avenue, Tucson, AZ 85719, USA; shaw@noao.edu, letizia@noao.edu}
\altaffiltext{7}{Istituto Nazionale di Astrofisica - Osservatorio Astrofisico di
Catania, Via S. Sofia 78, 95123 Catania, Italy; franco.cataldo@fastwebnet.it}
\altaffiltext{8}{Actinium Chemical Research, Via Casilina 1626/A, 00133 Rome,
Italy}


\begin{abstract}
We present a study of sixteen Planetary Nebulae (PNe) where fullerenes have been
detected in their {\it Spitzer Space Telescope} spectra. This large sample of
objects offers an unique opportunity to test conditions of fullerene formation
and survival under different metallicity environments as we are analyzing five
sources in our own Galaxy, four in the Large Magellanic Cloud, and seven in the
Small Magellanic Cloud. Among the sixteen PNe under study, we present the first
detection of C$_{60}$ (possibly also C$_{70}$) fullerenes in the PN M 1-60 as
well as of the unusual $\sim$6.6, 9.8, and 20 $\mu$m features (attributed to
possible planar C$_{24}$) in the PN K 3-54. Although selection effects in the
original samples of PNe observed with {\it Spitzer} may play a potentially
significant role in the statistics, we find that the detection rate of
fullerenes in C-rich PNe increases with decreasing metallicity ($\sim$5\% in the
Galaxy, $\sim$20\% in the LMC, and $\sim$44 \% in the SMC) and we interpret this
as a possible consequence of the limited dust processing occuring in Magellanic
Cloud  (MC) PNe. {\it CLOUDY} photoionization modeling matches the observed IR
fluxes with central stars that display a rather narrow range in effective
temperature ($\sim$30,000$-$45,000 K), suggesting a common evolutionary status
of the objects and similar fullerene formation conditions. Furthermore, the data
suggest that fullerene PNe likely evolve from low-mass progenitors and are
usually of low-excitation. We do not find a metallicity dependence on the
estimated fullerene abundances. The observed C$_{60}$ intensity ratios in the
Galactic sources confirm our previous finding in the MCs that the fullerene
emission is not excited by the UV radiation from the central star. {\it CLOUDY}
models also show that line- and wind-blanketed model atmospheres can explain
many of the observed [NeIII]/[NeII] ratios by photoionization suggesting that
possibly the UV radiation from the central star, and not shocks,  are triggering
the decomposition of the circumstellar dust grains.  With the data at hand, we
suggest that the most likely explanation for the formation of fullerenes and
graphene precursors in PNe is that these molecular species are built from the
photo-chemical processing of a carbonaceous compound with a mixture of aromatic
and aliphatic structures similar to that of HAC dust.
\end{abstract}





\section{INTRODUCTION}
The existence of fullerenes, highly resistant molecules formed exclusively by
carbon atoms in a number of 60 and more, was predicted in the 70s as a possible
variety of carbon assemblage in the Universe (Osawa 1970). However, it was only
in the 80's when the most common fullerenes C$_{60}$ (buckyball) and C$_{70}$
were synthetized in the laboratory for the first time (Kroto et al. 1985) in an
effort to trace the carriers of the elusive diffuse interstellar bands (DIBs).
Since then, they have been found on Earth (e.g., Buseck et al. 1992), on
meteorites (e.g., Becker et al. 1999)\footnote{Note that the fullerene
detections on Earth and meteorites are somewhat controversial and it is not
firmly sure that these authors have really found fullerenes in their matrices.},
in Planetary Nebulae (Cami et al. 2010; Garc\'{\i}a-Hern\'andez et al. 2010,
2011), and in the interstellar medium (Sellgren et al. 2010). More recently,
fullerenes have also been detected in a proto-planetary nebula (Zhang \& Kwok
2011), around R Coronae Borealis (RCB) stars (Garc\'{\i}a-Hern\'andez, Rao \&
Lambert 2011a,b), in the Orion nebula (Rubin et al. 2011), in post-AGB
circumbinary discs (Gielen et al. 2011), young stellar objects (YSOs), and
around a Herbig Ae/Be star (Roberts, Smith \& Sarre 2012).

From the variety of environments were fullerene molecules have been recently
detected the picture is emerging that the conditions for fullerene formation
cannot be too stringent since they have to be shared by star forming regions,
evolved stars, and the interstellar medium (ISM). In that sense, the detection
by the {\it Spitzer Space Telescope} of fullerenes in a remarkably
large/considerable number of Planetary Nebulae (16 sources so far)\footnote{Note
that instead, fullerenes have been detected towards a few - sometimes even only
one - targets in other astrophysical environments such as RCB stars, YSOs, etc.}
offers a unique opportunity to test the conditions of fullerene formation and
survival. 

Planetary Nebulae (PNe) represent the evolutionary stage during which the slow,
intense dust-driven winds, which are the signature of the late Asymptotic Giant
Branch (AGB) evolution, become ionized. During the PN phase, the effective
temperature of the central star can reach 200,000 K  while a fast, low density
post-AGB wind, shocks and shapes the inner parts of the previously ejected AGB
envelope. Chemically, the envelope shredded from the AGB star can show a C-rich
or O-rich chemistry, primarily as a direct consequence of the number and
efficiency of dredge-up processes experienced by the central star. These
processes ultimately are just determined by the stellar initial mass and its
evolutionary stage (Herwig 2005). Dynamically, the evolution of the shocked
ejected shell is mainly determined by both the stellar evolutionary stage and
core mass (see e.g., Villaver, Manchado \& Garc\'{\i}a-Segura 2002; Villaver,
Garc\'{\i}a-Segura \& Manchado 2002). In addition, during the PN phase, two
fundamental chemical drivers, i.e.,  the UV radiation field from the central
star and the shocks generated by the stellar winds, are present and quickly
evolving (see e.g., Kwok 2004).

From laboratory experiments we can learn a lot on the formation process of
fullerenes (Kroto et al. 1985; Kratschmer et al. 1990; de Vries et al. 1993)
and, in fact, the first synthesis of fullerenes in the lab motivated the first
searches in space. However, they can also be misleading. Although the laboratory
synthesis of fullerenes shows that they are more efficiently produced under
hydrogen-deficient conditions, this does  not seem to be the case in space,
where fullerenes seem to be efficiently formed in H-rich circumstellar
environments (Garc\'{\i}a-Hern\'andez et al. 2010, 2011). This is manifested by
the presence of C$_{60}$ fullerene bands in {\it Spitzer} spectra of Galactic
and extra-Galactic PNe with normal H-rich central stars that also display the
classical aromatic infrared bands (AIBs) usually attributed to polycyclic
aromatic hydrocarbons (PAHs) (Garc\'{\i}a-Hern\'andez et al. 2010, 2011) and
that includes the PN Tc 1.  Interestingly, fullerenes have been detected in PNe
whose IR spectra are clearly dominated by aliphatic carbon-rich dust,
represented by broad emissions such as those centered at $\sim$6$-$9, 9$-$13,
15$-$20, and 25$-$35 $\mu$m\footnote{Note that the 15$-$20 $\mu$m emission may
be also related with the C-C-C bending modes of relatively large PAHs or PAH
clusters (e.g., Van Kerckhoven et al. 2000).}. The different spectral properties
of these broad features observed from source to source are quite consistent with
the expected different properties of the hydrogenated amorphous carbon (HAC)
grains (or their decompositon products), which may be present in the
circumstellar envelopes of these stars, and that would explain the wide range of
different spectra observed depending on their physical and chemical properties
(e.g., size, shape, degree of hydrogenation; see Garc\'{\i}a-Hern\'andez et al.
2010, 2011). This raises the exciting possibility that considerable physical and
chemical processing of dust occurs just after the AGB phase (see also Zhang \&
Kwok 2011), including the formation of fullerenes. Moreover, fullerenes, being
relatively hardy molecules, may survive indefinitely in space playing an
important role in circumstellar/interstellar chemistry and physics.

From the experimental point of view, only very recently, 
laboratory spectroscopy of C$_{60}$ and C$_{70}$ fullerenes
(Iglesias-Groth et al. 2011) has provided the temperature dependence of
wavelength shifts and molar absorptivity of the C$_{60}$ and C$_{70}$ infrared
bands, which are fundamental for a firm identification of the observed bands as
fullerenes and for a quantitative abundance determination of these molecules in
space.

In this paper, we present a study of the infrared properties of sixteen
fullerene PNe, including  the first detection of fullerenes in
the Galactic PN M 1$-$60 and of the unusual $\sim$6.6, 9.8, and 20 $\mu$m
features (attributed to possible planar C$_{24}$) in the PN K 3-54. Our sample,
represents the largest (irrespective of the object type) in which fullerene
molecules have been detected, and includes for the first time objects belonging
to very different metallicity environments (the Galaxy and the Large and Small
Magellanic Clouds), allowing the study of how metallicity influences the chemical
route for fullerene formation. The {\it Spitzer} infrared data used in our study
are described in Section 2 while the data analysis, including the
estimation of C$_{60}$ and C$_{70}$ fullerene abundances, is presented in
Section 3. Section 4 discusses the results of our study and we offer the
conclusions of this work in Section 5.

\section{THE DATA}
We have searched for the presence of the infrared features atributed to the
C$_{60}$ and C$_{70}$ molecules in the spectra of a large sample of PNe (mainly
from our own {\it Spitzer} programs) and available in the {\it Spitzer/IRS}
archive. This includes the infrared spectra obtained under {\it Spitzer} GO
programs  \# 3633 (P.I.: M. Bobrowsky), and \# 20443 and \# 50261 (P.I.: L.
Stanghellini). A total of 238 {\it Spitzer/IRS} spectra of PNe were obtained
under these programs  with the general goal of investigating the chemical
properties and dust evolution during this evolutionary stage and
has been published elsewhere in studies dealing
with the original goals as proposed to the {\it Spitzer} observatory. All {\it
Spitzer} programs have in common the spectral coverage in the $\sim$5$-$38
$\mu$m range, making use of different combinations of the Short-Low (SL:
5.2$-$14.5 $\mu$m; 64 $<$ R$ <$ 128), Long Low (SL: 14.0$-$38 $\mu$m; 64 $<$ R
$<$ 128), Short-High (SH: 9.9$-$19.6 $\mu$m; R$\sim$600) and Long-High (LH:
18.7$-$37.2 $\mu$m; R$\sim$600) modules depending on the source brightness in
the mid-IR\footnote{Note that a minimum S/N of $\geq$50 is usually reached.}.
Detailed descriptions of observations and the data reduction process can be
found in Perea-Calderon et al. (2009) (program \# 3633) for the 40 PNe observed
in the Galactic Bulge, in Stanghellini et al. (2007)  and Shaw et al. (2010)
(program \# 20443) for the 41 PNe in the Magellanic Clouds (MCs), and in
Stanghellini et al. (2012) (program \# 50261) for the 157 compact PNe observed
in the Galactic Disk. An additional sample of 25 PNe in the MCs were observed
under  GTO program \# 103 (see Bernard-Salas et al. 2009) and we have included
these data in the analysis as well.  In summary, a total of 263 PNe have been
used in this paper in the search for fullerene emission and in the  analysis of
the characteristics of its emission that follows.

\subsection{{\it Spitzer/IRS} Spectra of PNe with Fullerenes}
The C$_{60}$ fullerene molecule displays four infrared features located at
$\sim$7.0, 8.5, 17.4, and 18.9 $\mu$m (e.g., Kratschmer et al. 1990;
Iglesias-Groth et al. 2011). In order to better isolate  these spectral features
we have performed a careful substraction of the dust continuum emission between
5 and 25 $\mu$m in the spectra of the 263 PNe mentioned above. To do so, the dust continuum
emission was represented by 3-7 order polynomials fitted at spectral locations
that we are confident to be free from any dust or gas feature; e.g., the
spectral points at 6, 9, 14, and 24 $\mu$m are not affected by dust emission at
6$-$9, 9$-$13, 15$-$20, and 25$-$35 $\mu$m. Note that we
prefer to exclude the 25-38 $\mu$m spectral region from the dust continuum
subtraction because the 25-35 $\mu$m emission sometimes extend well beyond 38
$\mu$m, difficulting the location of the continuum. Then, by inspecting the dust
continuum subtracted or residual spectra, we search for emission features at the four
spectral positions corresponding to the bands that may be
attributed to C$_{60}$ fullerenes (see Figure 1). To illustrate the technique in
Figure 1 we show examples of the dust continuum substracted spectra of the
PNe Tc 1 and K 3-54. Tc 1 (see left panel in Figure 1) was
observed at high-resolution and displays some of the strongest fullerene
features in our sample. On the other hand, PN K 3-54 (see right panel in Figure
1) was observed at low resolution and displays much weaker fullerene
bands than Tc 1.

We found obvious fullerene features in 16 objects out of the 263 PNe analyzed.
They are distributed as follows: seven SMC PNe (six from program \# 20443 and
one from program \# 103), four LMC PNe (three from program \# 20443 and one from
program \# 103), and five Galactic PNe (three from program GO \# 3633 and two of
them from GO \# 50261). Most of these detections have already been published: 3
Galactic and one SMC targets in Garc\'{\i}a-Hern\'andez et al. (2010); the ten
detections of fullerenes in PNe of the Magellanic Clouds in
Garc\'{\i}a-Hern\'andez et al. (2011); and the first fullerene detection in Tc 1
in Cami et al. (2010). In this paper, we report the detection of fullerenes in a
new source, the Galactic PN M 1-60, and we provide an overall and detailed
analysis of the whole sample of PNe with fullerenes.

\section{ANALYSIS}
Figures 2 to 4 display the {\it Spitzer/IRS} spectra of all fullerene
PNe discovered so far grouped by location or metallicity environment, that is,
we show in Fig 2 the targets in the Galaxy while Figs. 3 and 4 show the PNe in
the LMC and SMC, respectively. Note that in these figures the strongest C$_{60}$
feature at 18.9 $\mu$m (sometimes also C$_{60}$ 17.4 $\mu$m) is distinctly
present superimposed on the dust continuum thermal emission.   Most of the PNe
also show broad dust emission features centered at $\sim$9$-$13 and/or 25$-$35
$\mu$m (the so-called 30 $\mu$m feature; e.g., Hrivnak, Volk \& Kwok 2000). All
C$_{60}$-detected PNe exhibit either the broad 30 $\mu$m, or the 9$-$13 $\mu$m
feature, but if not, then they display a very prominent 6$-$9 $\mu$m feature,
like SMC 16 and SMC 24 (Figure 4) where this feature is the strongest among the
three features mentioned above. A huge 6$-$9 $\mu$m feature is also seen in LMC
02 (Fig. 3), which does not show any 9$-$13 nor 30 $\mu$m features. SMC 20 (Fig.
4) is a peculiar case because it shows the strongest 9$-$13 and 15$-$20 $\mu$m
features together with the complete lack of the 30 $\mu$m feature. 

M 1-60, a new fullerene PN\footnote{M 1-60 also displays the weak
AIBs at 6.2, 7.7, and 11.3 $\mu$m usually attributed to PAHs.}, displays an
extremely weak 9$-$13 $\mu$m feature together with a wider and sub-structured 30
$\mu$m feature -  narrower (FWHM$\sim$0.5-3.5 $\mu$m) emission features
superimposed to the broad 30 $\mu$m emission are seen at $\sim$25.9, 27.8, 29.7,
and 34.1 $\mu$m. The 30 $\mu$m feature seen in M 1-60 resembles that of the PPN
IRAS 22574$+$6609 (Hrivnak, Volk \& Kwok 2000), which also shows a narrow and
unidentified emission feature at 25.8 $\mu$m. Interestingly, a similar pattern
(although the features look weaker and narrower) of emission sub-features is
also seen in the other Galactic fullerene PNe such as Tc 1, M 1-20,
and M 1-12 (Figure 2) for which high-resolution (R$\sim$600) {\it Spitzer/IRS}
spectra are available. We believe that these unidentified 30 $\mu$m sub-features
are possibly related with the formation of fullerenes, in the sense that their
carriers may be decomposition products of HACs (or a similar carbonaceus
compound like coal, petroleum fractions, etc.) such as fullerene precursors. 

It is worth to note here that the exact positions (and widths) of the $\sim$7.0,
8.5, 17.4, and 18.9 $\mu$m C$_{60}$ features are temperature dependent (see
Iglesias-Groth et al. 2011 and references therein). Duley \& Hu (2012) have very
recently suggested that these four IR features may be due to proto-fullerenes
rather than C$_{60}$ in objects that also show the 16.4 $\mu$m feature and other
AIBs usually attributed to PAHs. The 16.4 $\mu$m feature is usually seen when
the classical AIBs at 6.2, 7.7, and 11.3 $\mu$m are much stronger than the
$\sim$7.0, 8.5, 17.4, and 18.9 $\mu$m IR features, being the case for sources
like reflection nebulae (NGC 7023, NGC 2023), RCB stars (DY Cen, V854 Cen),  the
proto-PN IRAS 01005$+$7910 and a few post-AGB stars (Gielen et al. 2011). Thus,
it is possible that the $\sim$7.0, 8.5, 17.4, and 18.9 $\mu$m IR features are
due to proto-fullerenes in these sources (Garc\'{\i}a-Hern\'andez et al. 2012).
However, the PN Tc 1 does not display strong AIBs nor the 16.4 $\mu$m feature,
being a proxy for a source where only C$_{60}$ fullerenes are detected.  The
16.4 $\mu$m emission feature is not present either in any of the PN sources
analyzed in this paper, and the AIBs are very weak--contrary to the case in
reflection nebulae, post-AGB stars, and RCB stars-- and therefore the most
reasonable assumption is that the $\sim$7.0, 8.5, 17.4, and 18.9 $\mu$m emission
features are mostly due to C$_{60}$ fullerenes. 

Some of the PNe in our sample (e.g., SMC 16, M 1-12, K 3-54, SMC 24, LMC 02) are
almost spectroscopic twins of Tc 1, showing the broad 6$-$9 $\mu$m emission
plateau and the 9$-$13, and 30 $\mu$m dust emission features characteristic of
aliphatic hydrocarbons (see e.g., Kwok \& Zhang 2011) such as HAC aggregates or
very small grains (VSG) (e.g., Tielens 2008). Thus, they should be in an
evolutionary stage similar to Tc 1.  This interpretation is supported by the
narrow range of central star effective temperatures that all these sources have
(see \S 3.1).

It is also important to consider that we lack the knowledge of the relative
spatial distribution of the carrier of the AIBs with respect to C$_{60}$ emission, 
which is an important aspect of the analysis and that could offer important
clues on the formation mechanism of fullerenes in PNe. The PNe are spatially
small and the {\it Spitzer} spectrum generally encompases the whole nebula.  Tc
1 is the only  exception, being the most extended source in our sample. In Tc 1
the fullerene and PAH emission seem to peak slightly offset from the central
star (where the dust continuum emission peaks) and they look more extended by
1-2 pixels than the dust continuum emission. Indeed, a careful inspection of the
Tc 1's {\it Spitzer} spectrum reveals the presence of very weak PAH 6.2 and 11.3
$\mu$m emission; see e.g., Figures 2 and 3 in Garc\'{\i}a-Hern\'andez et al.
(2010). Unfortunately, the {\it Spitzer/IRS} observations contain a marginal
amount of information at a spatial resolution of $\sim$2". We believe
that mid-IR images at much higher spatial resolution are needed in order to
reach any firm conclusion about the relative spatial distribution of C$_{60}$,
AIBs, and other emission in Tc 1.

Figure 5 displays the residual spectra of the galactic fullerene PNe Tc 1, M
1-20, M 1-12, and K 3-54 and the extragalactic PN LMC 02. The expected positions
of the four  C$_{60}$ bands are marked with dashed vertical lines. We would like
to remark that two additional MC PNe (SMC 17 and SMC 19; see Stanghellini et al.
2007 for the spectra) show a strong feature at 19.0 $\mu$m that could be
attributable to C$_{60}$. However, these PNe show very strong AIBs (e.g.,
those at 6.2, 7.7, 8.6, and 11.3 $\mu$m usually attributed to PAHs) together
with a yet unidentified feature at 17.0 $\mu$m (see e.g., Boersma et al. 2010).
Given that  the other C$_{60}$ features (at $\sim$7.0 and 17.4 $\mu$m) are not
obvious in their residual spectra, and that the 17.0 and 19.0 $\mu$m features
may be due to very large PAHs (Boersma et al. 2010) we considered that these two
sources do not have solid fullerene detections, and therefore they have not been
included in the analysis presented here.

For three (Tc 1, LMC 02, and SMC 24) of the C$_{60}$-containing PNe, high
resolution (R$\sim$600) {\it Spitzer/IRS} spectra are also available, in which
relatively strong and isolated C$_{70}$ features are present at 12.6, 14.9,
15.6, 17.8, 18.7, and 21.8 $\mu$m (von Czarnowski \& Meiwes-Broer 1995; see
Figure 2 in Garc\'{\i}a-Hern\'andez et al. 2011). The C$_{70}$ infrared features
may be present in PNe other than Tc 1, LMC 02, and SMC 24. Only low resolution
{\it Spitzer/IRS} spectra were obtained for the MC PNe in our sample, avoiding
us to confirm/exclude the detection of these features in more extra-galactic
sources. However, the Galactic PNe M 1-20, M 1-12, and M 1-60 were observed at
high resolution as well and these three PNe show a weak but unblended feature at
21.8 $\mu$m (see Figure 2) that we attribute to C$_{70}$. Consistently with our
identification, M 1-60 and M 1-20 show also some C$_{70}$ emission at 14.9
$\mu$m (Figure 5). We do not estimate the C$_{70}$ abundances and temperatures
in these three sources due to the intrinsic weakness of these possible C$_{70}$
features and because we only see a maximum of two unblended C$_{70}$ features
(see Section 3.2).

We report for the first time the presence of unusual IR features at $\sim$6.6,
9.8, and 20 $\mu$m, which are coincident with the strongest transitions of
planar C$_{24}$ (graphene precursors or ``proto-graphene") in the Galactic PN K
3-54. These features have been detected in another two targets that also have
fullerenes, LMC 02 and SMC 24 (Garc\'{\i}a-Hern\'andez et al. 2011).  Among the
other Galactic fullerene PNe, M 1-12 also displays a strong 6.6
$\mu$m feature but not the other ones. Although more tentative, the 6.6 $\mu$m
feature seems to be present also in Tc 1, M 1-20, and M 1-60 (see Figure 5). 

In Table 1 we give the positions and integrated fluxes for the C$_{60}$ and
C$_{70}$ fullerene bands identified in the 16 fullerene PNe, as well as the
possible planar C$_{24}$ features. These measurements  have been done in the
residual spectra by fitting a baseline to the local continuum of each infrared
band in order to subtract any possible extra-contribution from the underlying
dust features at 6$-$9, 9$-$13, and 15$-$20 $\mu$m (see Figures 2 to 4) still
present in the dust continuum subtracted spectra. We also list in Table 1 our
estimated flux errors between brackets, which take into account the error in the
measurement as well as the error introduced by the uncertainty of the
determination of the local baseline selected for integrating the fluxes. Our
estimated flux errors are typically $\sim$15\% for the Galactic PNe as well as
for the strongest fullerene transitions (e.g., at 17.4 and 18.9 $\mu$m) in the
PNe of the Magellanic Clouds. The estimated flux errors are, however, higher for
the weaker fullerene transitions (e.g., at 7.0 and 8.5 $\mu$m) observed in
Magellanic Cloud PNe, which sometimes can be as high as $\sim$30-40\% (Table 1).
Note also that the observed 7.0 $\mu$m emission is a blend of C$_{60}$,
C$_{70}$, and [Ar II] 6.99 $\mu$m emission. We estimate the C$_{60}$ and
C$_{70}$ 7.0 $\mu$m flux contributions from the corresponding Boltzmann
excitation diagrams below (\S 3.2). The position and flux for the 15.6 $\mu$m
C$_{70}$ band is only given when this feature is not blended with the usually
stronger 15.6 $\mu$m [Ne III] emission. 

Table 2 gives some basic physical parameters when available in the literature
such as the effective temperature of the central star, the H$_{\beta}$ flux,
carbon abundance, electron temperature and density, and total hydrogen and
carbon mass of the nebula.

\subsection{{\it CLOUDY} Photoionizaton Models with Line- and Wind-blanketed Atmospheres}
In a previous work we presented some basic {\it CLOUDY} models (Ferland et al.
1998) for the fullerene PNe in the Magellanic Clouds (Garc\'{\i}a-Hern\'andez et
al. 2011). We were only interested in doing some general and basic model
predictions to explore the possible origin of the 7.0 $\mu$m emission as well as
of the usually low [Ne III]/[Ne II] ratios observed in fullerene PNe. To do so
we adopted for the stellar ionizing continuum blackbodies (BB) and stellar
spectra taken from Rauch (2003) and concluded that the 7.0 $\mu$m emission in
fullerene PNe cannot be attributed solely to [Ar II] 6.99 $\mu$m but to a
combination of C$_{60}$ and C$_{70}$. Furthermore, the fact that those models
were unable of explaining the low [Ne III]/[Ne II] ratios by photoionization,
and given that fast stellar winds from the central stars are ubiquitous in PNe
(and the associated shocks) lead us to the suggestion of a possible
shock-excited origin for the [Ne II] 12.8 $\mu$m emission in fullerene PNe (see
Garc\'{\i}a-Hern\'andez et al. 2011 for more details). We did not contemplate
the possibility, however, that the observed [Ne III]/[Ne II] ratios might be
also explained by using model atmospheres with line (and wind) blanketing (C.
Morisset 2011, private communication; see also Morisset et al. 2004 and
references therein). Here we present new radiation transfer modeling exploring
this option.

We have used version C08.01 of the photoionization code {\it CLOUDY} to model
the nebular emission in our fullerene sources. For the model atmosphere of the
central star, this time we have used the grid of models provided by Pauldrach et
al. (2001), which are available in {\it CLOUDY}, identified as tables WMBasic. These
models include non-LTE, line- and wind-blanketed model atmospheres for O-B stars
with solar and sub-solar metallicities. For all models we have assumed a
constant luminosity of $10^4$~L$_\odot$ of the star and a fixed distance to the
nebula of 0.1 pc (inner radius).  Variations of luminosity and/or inner radius
closely mimics variations of the temperature of the central star. Thus, our
sequences (varying central star temperature) represents the locus of a given
model for any combination of luminosity, distance and temperature. We have
computed models using solar and sub-solar metallicities for the atmosphere of
the central star, but there are no large differences in the resulting line
ratios. We have assumed for the abundances of the nebula those for the prototype
fullerene PN Tc 1 and quoted by Pottasch et al. (2011) in Table 15. We have
considered a spherical geometry for the nebula with a density of 1800 cm$^{-3}$
at the inner radius, and a dependency with the distance to central ionizing
source as r$^{-2}$.  The chemical abundances of the photoionized gas are also
those quoted by Pottasch et al. (2011) in their Table 15. 

In Figure 6 we display our model predictions compared to the observed line
ratios ([ArII]/HI vs. [ArIII]/HI and [SIV]/[SIII] vs. [NeIII]/[NeII]) in the
fullerene PNe of our sample. In the right panel we present a
combination of line ratios ([SIV]/[SIII] vs. [NeIII]/[NeII]), which permits to
determine the effective temperature (T$_{eff}$) of the central star,
independently of the nebular abundances. It can be clearly seen that model
atmospheres with line and wind blanketing can explain many of the observed
[NeIII]/[NeII] ratios. In addition, all fullerene sources show a rather narrow
range in effective temperature (T$_{eff}$$\sim$30,000$-$45,000 K). 

The fact that line and wind blanketing can explain many of the observed
[NeIII]/[NeII] ratios weakens the hypothesis of a shock-excited origin for the
[Ne II] 12.8 $\mu$m emission. Unfortunately, the shock hypothesis cannot be
directly tested with the present data.

It is to be noted here that the observed central star effective temperatures
determinations in the literature (those listed in Table 2) came mostly from
hydrogen Zanstra analysis, and thus although more reliable for optically thick
PNe, they should always be used as lower limits for optically thin targets (see
also \S 4.1). Moreover, none of the temperature determinations available are
free from model assumptions. Indeed, from our {\it CLOUDY} modeling (see Figure 6) we
infer effective temperatures higher than the hydrogen Zanstra temperatures
(although always lower than 45,000 K) for all of these MC PNe.  The only
exception is LMC 02 for which the [S IV]/[S III] ratio is somewhat lower than
expected for its T$_{eff}$ (39,000 K as derived from FUSE data); we atribute
this difference to a contamination by C$_{70}$ of the observed 18.7 $\mu$m
emission. From Figure 6 is also evident that for Galactic PNe M 1-20 and M 1-60
the models indicate T$_{eff}$ $\sim$ 40,000-45,000 K\footnote{The {\it CLOUDY}
T$_{eff}$  for M 1-20 and M 1-60 are significantly lower thant the literature
values (Table 2) and derived from the Zanstra HI method in M 1-20 (Kaler \&
Jacoby 1991) and from the Energy-Balance method in M 1-60 (Preite-Mart\'{\i}nez
et al. 1989).}, whereas for Tc 1 T$_{eff}$ $\simeq$ 35,000 K. Note that the
Galactic PNe M 1-12 and K 3-54 are not shown in this diagram because they only
present [Ne II] 12.8 $\mu$m and [S IV] 10.5 $\mu$m, emission, respectively. An
anomalous high value of [Ne II] 12.8 $\mu$m emission (e.g., as a consequence of
strong shocks) in these sources cannot be discarded.
  
In the left panel of Figure 6 ([ArII]/HI vs. [ArIII]/HI) we illustrate how varying
the abundance of Ar affects noticeably the  line ratios involving these lines. It
can be seen that the observed values of [Ar III]/H I can be explained with the
model effective temperature, consistently with the values derived from the
[SIV]/[SIII] vs. [NeIII]/[NeII] diagram shown in the right panel of Figure 6.
However, none of the models can explain (even taking into account maximum errors of
40\% in the observed line intensity ratios) the high observed values of the [Ar
II]6.99 $\mu$m/HI line flux ratio. For example, in the case of Tc 1 and M 1-12, the
expected ratios are around 1, whereas in the case of M 1-20 and M 1-60 those are
around 0.1. Note that PN K 3-54 does not appear in this diagram because the HI 7.5
$\mu$m line is blended with the much stronger 7.7$\mu$m AIB. Thus, the new {\it
CLOUDY} modeling confirms that the 7.0 $\mu$m emission in fullerene  PNe cannot be
attributed uniquely to [Ar II] 6.99 $\mu$m and the C$_{60}$ and C$_{70}$ emission
contributions have to be taken into account.

\subsection{Boltzmann Excitation Diagrams of the Fullerene Emission}

Our previous work indicates that the fullerene emission in PNe of the Magellanic
Clouds is likely originated from solid-phase fullerene molecules, being formed
on the surface of pre-existing very small grains (or HACs)
(Garc\'{\i}a-Hern\'andez et al. 2011; Duley \& Hu 2012). This conclusion was
based on the fact that the observed C$_{60}$ 7.0/18.9, 8.5/18.9, and 17.4/18.9
intensity ratios differ from the theoretical predictions for UV-excited
gas-phase C$_{60}$ molecules given by Sellgren et al. (2010). Here we have
determined the fluxes of the 7 and 8.5 $\mu$m C$_{60}$ transitions (which are
actually only due to C$_{60}$) in the Galactic PNe M 1-20, M 1-12, M 1-60, and K
3-54\footnote{Note that the observed 8.5 $\mu$m fluxes cannot be completely
explained by C$_{60}$ emission in a few MC PNe (SMC 13, 24, LMC 02, 99; see
Garc\'{\i}a-Hern\'andez et al. 2011) and in two Galactic PNe (K 3-54 and M
1-12) only. Thus, these objects may show some AIB emission (e.g., from
PAHs) at this wavelength.} and we have constructed the corresponding Boltzmann
excitation diagrams (see below for more details) for these Galactic PNe. 

In Table 3 we list the C$_{60}$ intensity ratios in fullerene PNe of our Galaxy
in comparison with the values derived for fullerene PNe in the Magellanic
Clouds. Note that the listed C$_{60}$ F(7.0)/F(18.9) flux ratios are obtained by
correcting the observed emission fluxes at 7 $\mu$m by the C$_{60}$ contribution
factors (f$_{C60}$) derived from the Boltzmann excitation diagrams (see below)
and also listed in Table 3. In addition, the listed C$_{60}$ F(8.5)/F(18.9)
intensity ratios in a few sample PNe (see above) were corrected by their
corresponding contribution factors at 8.5$\mu$m (not shown in Table 3). From the
results obtained we can reach a similar conclusion about the fullerene
excitation mechanism for the subsample of fullerene-detected PNe in our Galaxy
(see Table 3). In particular, the C$_{60}$ 7.0/18.9 and 8.5/18.9 ratios indicate
photon energies of less than 5 eV, regardless of the central star temperatures
(between $\sim$30,000 K and 45,000 K). In addition, the C$_{60}$ 17.4/18.9
intensity ratio strongly varies from source to source (from 0.21 to 1.13), while
this ratio is predicted to be unsensitive to the energy of the absorbed UV
photons (a C$_{60}$ 17.4/18.9 ratio of 0.28$-$0.38 is predicted by Sellgren et
al. 2010) for gas-phase fullerene molecules. It should be noted here that our
conclusion is different from that very recently drawn by Bernard-Salas et al.
(2012). However, the other fullerene C$_{70}$ also contributes to the observed 7
$\mu$m flux (Iglesias-Groth et al. 2011; Garc\'{\i}a-Hern\'andez et al. 2011) -
e.g., in Tc 1 the C$_{70}$ flux contribution (40\%) is even higher than that for
C$_{60}$ (25\%). Our conclusion is supported by the fact that the same
temperatures (within the errors) are needed to explain both the 7.0 and 8.5
$\mu$m C$_{60}$ band strengths. Finally, we note that taking into account the
uncertainties of flux measurements and Einstein-coefficients, our average
intensity C$_{60}$ F(7.0)/F(18.9) (=0.28), F(8.5)/F(18.9) (=0.35), and
F(17.4)/F(18.9) (=0.56) ratios would be more consistent with the C$_{60}$
intrinsic strengths reported by Fabian (1996) and Iglesias-Groth et al. (2011).
Thus, in the following we will assume that the fullerene emission in PNe is
originated from solid-state fullerenes. 

We can determine the fullerene excitation temperature for each PN from the flux
measurements of the C$_{60}$ and C$_{70}$ transitions (see Table 1) and the
absorptivity measurements recently obtained by Iglesias-Groth et al. (2011).
Absorptivity for each transition led to an f-value from which the Einstein
coefficient A$_{ij}$, is derived which provides a relation between the emitted
flux and the number of molecules populating the energy level associated with the
transition (see e.g., Cami et al. 2010 for more details).

We adopted the temperature dependence of the absorptivity for each C$_{60}$ and
C$_{70}$ transition given in Iglesias-Groth et al. (2011) and developed an
iterative process to find the best possible correlation coefficient in the plane
ln(N$_{u}$/g$_{u}$) vs. E$_{u}$/k independently for each fullerene. A thermal
distribution over the excited states will be indicated by colinear datapoints in
the Boltzmann excitation diagram. In this case, the slope and the intercept of
the linear relation determines the temperature and the total number of emitting
fullerene molecules. The initial temperatures for this iterative process were
assumed in the range from 100 to 1000 K and only a few iterations were required
for a quick convergence. The temperature values for C$_{60}$ and C$_{70}$
derived this way are listed in Table 3. The four mid-IR transitions of C$_{60}$
were considered, allowing the fraction of the flux of the 7$\mu$m transition -
which is actually due to C$_{60}$ - as a free parameter because of the possible
contamination with C$_{70}$ and the [Ar II] nebular line. The best linear fit is
sensitive to this fraction, and thus we also iterated on this parameter in order
to explore the full range of possible values (from 0 to 1). In the case of
C$_{70}$, we considered the 3 or 4 (depending on the source and the S/N of the
spectrum; see Table 1 for the C$_{70}$ features used for each source)
mid-infrared transitions less contaminated by other species, iterating also on
the fraction of the 7$\mu$m flux that may be due to C$_{70}$. We generally found
very good correlation coefficients that are always higher than 0.89. Figure 7
displays illustrative examples of the Boltzmann excitation diagrams for the
C$_{60}$ (left panel) and C$_{70}$ (right panel) fullerene emission detected in
the MC PNe SMC 16 and LMC 02, respectively. 

The fullerene PNe in the Magellanic Clouds were previously studied by
us in Garc\'{\i}a-Hern\'andez et al. (2011). The sources in the MCs offer a more
reliable estimation of the C$_{60}$ and C$_{70}$ abundances in PNe than in their
Galactic counterparts. This is because the distances to the MCs (48.5 and 61 kpc
for the LMC and SMC, respectively) are accurately known and reliable C
abundances are also available in the literature (Garc\'{\i}a-Hern\'andez et al.
2011). However, this is not the case for the five fullerene PNe in
our Galaxy - in addition, only Tc 1 has a reliable C abundance in the literature
(Table 2). For consistency, we decided to assume the recent statistical
distances by Stanghellini \& Haywood (2010) for four of the Galactic sources: M
1-20 (8.0 kpc), M 1-12 (9.2 kpc), M 1-60 (9.5 kpc), and K 3-54 (23.4 kpc). The
only exception is Tc 1\footnote{Stanghellini \& Haywood (2010) give a
statistical distance of 2.7 kpc towards Tc 1.}, for which we assumed a distance
of 2 kpc for comparison with Cami et al. (2010). Following Milanova \& Kholtygin
(2009) we may assign a PN type (i.e., type I, IIa, IIb, III, IV)\footnote{For
type I ($|$z$|$= 0.23 kpc) A(C)=8.32, for type IIa ($|$z$|$= 0.31) A(C)=8.82,
for type IIb ($|$z$|$= 0.56) A(C)=8.55, for type III ($|$z$|$= 1.05) A(C)=8.60,
and type IV ($|$z$|$= 1.35) A(C)=8.64 (see Table 5 in Milanova \& Kholtygin
2009).}  and an average carbon abundance (A(C)=log(C/H)+12) to each Galactic PN
from the height above the Galactic plane z. Therefore we assume a type IIa and
A(C)=8.55 for M 1-20 (z=0.492 kpc) and K 3-54 (z=0.547 kpc) as well as type IIb
and A(C)=8.83 for M 1-12 (z=0.258 kpc) and M 1-60 (z=0.290 kpc). The IIab PN
types of the Galactic fullerene PNe, as derived by this very rough
and limited method, seem to be consistent with that of Tc 1, which is identified
as a slowly evolving low-mass type II PN (Garc\'{\i}a-Hern\'andez et al. 2010).
In addition, as we will see in Section 4.1, all fullerene PNe are
likely of low-mass and are slowly evolving towards the white dwarf stage. Table
2 lists all relevant information (e.g., H$_{\beta}$ flux, C abundance, electron
temperature and density, total hydrogen and carbon masses of the nebula) as well
as the source of information used in each case from the literature for all
fullerene PNe in our sample. 

By using the above distances, the number of C$_{60}$ and C$_{70}$ molecules (or
masses of pure C$_{60}$ and C$_{70}$) was estimated for all sources in our
sample and the C$_{60}$/C and C$_{70}$/C abundances listed in Table 3 were
derived from the total carbon and hydrogen mass of the nebula. The total H mass
of the nebula is obtained from the electronic temperature and density and the
observed H$_{\beta}$ flux, while the total carbon mass is obtained from the
total H mass and C abundance of the nebula. The corresponding C$_{70}$/C$_{60}$
ratios are also given in Table 3. We find C$_{60}$/C$\sim$0.01$-$0.05\% (average
of 0.03\%) and T$_{C60}$=260$-$660 K for the four Galactic fullerene
PNe studied for the first time in this paper. The formal error of our fullerene
abundance estimations are difficult to evaluate because we are using different
literature sources for the C abundance, electron density, etc. However, it
should be noted that by giving the C$_{60}$/C and C$_{70}$/C abundance ratios,
the distance dependence is canceled. Thus, the main sources of error are the
electronic density  and carbon abundance. We estimate that the errors of our
fullerene abundances in MC PNe are certainly lower than 50\% (likely much lower
than this value for most of the MC PNe sample) but the maximum error in the
C$_{60}$/C ratio for the Galactic PNe (with the exception of Tc 1 for which the C
abundance is more accurately known) may be a factor 2 or 3. Taking into account
all assumptions made in the Galactic objects, our individual and average
C$_{60}$/C abundances as well as the C$_{60}$ temperatures for these sources are
very similar to the C$_{60}$ abundances obtained in the MC ones (see Table 3).
Finally, note that we estimate C$_{60}$/C=0.04\% and C$_{70}$/C=0.005\% in Tc 1,
which are a factor of 40 and 300, respectively, lower than the $\sim$1.5\%
estimates for both species by Cami et al. (2010).  

\section{DISCUSSION}

\subsection{Evolutionary Status of the Fullerene PNe}

The temperature and luminosity of central stars of PNe are important parameters
in order to determine the evolutionary status of the nebula. Central star
temperatures are available  for a sizable number of objects in the sample of PNe
with fullerenes (9 PNe). However, most of the temperature  determinations of the
objects at hand rely on widely used indirect methods either Zanstra (Zanstra
1931), or the Energy Balance (Stoy 1933) with only one exception LMC 02 obtained
from FUSE data (Herald \& Bianchi 2004). Moreover, for those PNe for  which
Zanstra temperatures have been obtained (SMC 13, SMC 24, SMC 27 and LMC 25) only
hydrogen Zanstra temperatures could be determined for which the optically thick
nature of the nebulae to all the photons above the Lyman limit of H is not
assured. A hint to the fact that the effective temperature of these stars cannot
be much higher than $\approx$ 50,000 \,K is provided by the lack of detection of
He II 4686 \AA~nebular line flux and confirmed by our new {\it CLOUDY} modeling (see
Section 3.1).

Although not definitive conclusions can be drawn given the small number of
objects and the  uncertainties, the available data shows that all the PNe with
fullerenes have central stars with low effective temperatures in the range
between $\sim$30,000 and 45,000 K. Given the paths that these stars follow in
the HR diagram a low temperature can imply a very old or a very young star,
therefore a second parameter, stellar luminosity (or mass), is needed in order
to understand their  evolutionary status. Unfortunately, central star 
luminosity determinations rely on the availability of the distances which are
accurately  determined for a handful of objects in the Galaxy and none of the
fullerene containing PNe are among them. The situation is quite different for
the PNe located at the known distances of the Magellanic Clouds. In this case,
determinations of the central star luminosities are possible with great
accuracy  provided that the central star flux is measured. Out of the 11 PNe in
the clouds only 3 of them have meassured luminosities (see Villaver,
Stanghellini \& Shaw 2003, 2004); the reasons are either the objects have not
been observed with the  appropriate instrumentation in order to measure the
stellar flux (6 PNe), the star was saturated  (LMC 25), or the central star was
below the nebular level and thus not detected (SMC 13). 

PNe SMC 24, SMC 27, and LMC 02 are the only ones that we can analyze in full
extent given that they  have both luminosities and effective temperatures
available  (Villaver, Stanghellini \& Shaw  2004;  Herald \& Bianchi 2004).
Although conclusions from 3 objects cannot  be extrapolated to the full sample,
we find meaningful the fact that the 3 of them have luminosities that locate
them in the horizontal part of the post-AGB tracks for central stars of PNe.
Together with their low effective temperatures this indicate that, at least
these 3 objects are truly unevolved stars for  which less than $\approx$ 10,000
\,yr (central star time according to Vassiliadis \& Wood 1994) could have 
passed since the moment the star left the AGB phase. Moreover, all three objects
have central stars in the  low mass range: 0.59 M$_{\sun}$ for SMC 24; 0.60
M$_{\sun}$ for SMC 27, and 0.56 M$_{\sun}$ for LMC 02 implying that the three of
them descend from a low mass progenitor with a main sequence mass $<$ 1.5
M$_{\sun}$.  It is important to note that the masses of fullerene
systems in the SMC are within the average mass of central stars of PNe in the
SMC, 0.63 M$_{\sun}$ (Villaver, Stanghellini \& Shaw 2004), while  the LMC
object is significantly less massive than the average, 0.65 M$_{\sun}$ central
star mass in the  LMC (Villaver, Stanghellini \& Shaw 2003, 2007).

The central star and the nebular gas are interdependent systems, that is, the
evolution of the nebulae is governed by the energetics provided by the central
star mostly throught the stellar wind and the ionizing  radiation field.  In
that sense, the physical radius of the PNe can provide an indirect way of
evaluating  the evolutionary status of the systems provided that for the objects
in the MCs this quantity is available. If we take the photometric radius of the
nebulae\footnote{The photometric radius, R$_{phot}$,  corresponds to the size of
a circular aperture that contains 85\% of the flux in [O III] $\lambda$5007.
R$_{phot}$ gives an objective measurement of the nebular angular size that is
insensitive to the S/N ratio of the image and is useful for evolutionary
studies.} measured from {\it HST} data in Stanghellini et al. (2003) and Shaw et
al. (2006), then adopting a distance to the SMC of 61 kpc and to the LMC of 48.5
kpc we obtain the physical radius of the PNe in the MCs. With only one exception
LMC 99 (with a physical radius of 0.1 \,pc) all the radius are $<$ 0.06 \,pc.
The average radius of all the PNe in the MCs that  contain fullerenes is 0.058
\,pc (or 0.054 pc if we exclude LMC 99). The majority of the PNe in the {\it
HST} samples have small sizes; they were selected to avoid entangled
line in the STIS spectroscopic observations. They have a median  photometric radius
of 0.10 pc in the LMC and 0.08 pc in the SMC (see Shaw et al. 2006). The sample
of  MC PNe that contain fullerenes have physical sizes even smaller (most of them
$<$ 0.06 pc) than the average values. Irrespective of the mechanism for  shell
growth,  at the expense of the energy provided by the hot bubble (the high
velocity wind producing an adiabatic shock) or by the propagation of the
ionization front, it is expected to increase with time (see Villaver, Manchado
\& Garc{\'{\i}}a-Segura 2002). Thus, if we accept the physical size as an
indicator of the evolutionary status of the nebulae, then the
fullerene PNe in the MCs seem to be young  (non-evolved) systems as
suggested by their small physical radii; smaller than the average value of a
full sample. 

\subsection{Fullerene Detection vs. Metallicity}

Interestingly, the fraction of fullerene-detected PNe increases with decreasing
metallicity ($\sim$5\% in the Galactic Disk versus $\sim$31\% in the MCs). This
fraction goes from $\sim$5\% in the Galactic Disk (2 detections from 39 C-rich
compact ($<$4") and young PNe; Stanghellini et al. 2012) to $\sim$20\% in the
LMC (4 detections from 20 C-rich LMC PNe) and $\sim$44\% in the SMC (7
detections from 16 C-rich SMC PNe). In addition, the {\it Spitzer/IRS} spectra
reveal that the dust processing is very limited in the low metallicity
environments of the Magellanic Clouds. The size of the dust grains remains small
(e.g., Stanghellini et al. 2012) and the formation of PAHs (as indicated by the
presence of AIBs) is more difficult at low metallicities (Stanghellini et al.
2007). This indicates that the dust grains are less processed in low-metallicity
PNe than in their Galactic counterparts, likely because of the probability
of forming big carbon dust aggregates around the central star, which 
may be affected by the posterior PN evolution - i.e., the quickly changing UV
radiation field and/or the post-AGB shocks from the strong stellar winds. This
probability decreases with decreasing metallicity.

Although we find meaningful that probably the fraction of fullerene PNe depends
upon the metallicity of the host environment, it is worth to mention here that
selection effects in the {\it Spitzer} PN samples may play a potentially
significant role in the statistics mentioned above. None of these Magellanic
Cloud PN samples is complete, except that the surface brightness was high enough
to be included in a HST or {\it Spitzer} sample. However, our estimation of the
detection rate of fullerenes ($\sim$5\%) in our own Galaxy comes from the sample
of 157 compact and young PNe in the the Galactic Disk (Stanghellini et al.
2012). This Galactic fullerene detection rate may be considered as an upper
limit for environments of roughly solar metallicity. This is because the
Stanghellini et al. (2012)'s PNe sample is sampling the early stages of PN
evolution and dust processing (see below), when the detection rate of fullerenes
is expected to be maximum; the fullerenes are not too cold (e.g., T$_{C60,C70}$
$<$ 100 K) to escape detection at mid-infrared wavelengths.

In short, the high detection rate of fullerenes in MC PNe seems to be strongly
linked to the limited dust processing (or the general presence of small dust
grains) in low-metallicity circumstellar envelopes. This is consistent with the
observational finding that the still unidentified 21 $\mu$m feature (Kwok, Volk
\& Bernath 2001; see also Garc\'{\i}a-Hern\'andez 2012 for a recent review) is
much more common in post-AGB stars of the Magellanic Clouds than in our own
Galaxy (Volk et al. 2011), suggesting that the 21 $\mu$m feature is possibly also
related with the formation of fullerenes and its carrier may be a fragile
intermediate product from the decomposition of HAC or a similar material (see
below). 

\subsection{Broad 6$-$9, 9$-$13, 15$-$20, and 25$-$35 $\mu$m Emission}

As we have mentioned before, the dust processing seems to depend on metallicity.
For example, MC PNe display infrared spectra, which are similar to many Galactic
post-AGB stars where dust evolution is still in its early stages, and/or still
taking place. However, already in galactic post-AGB stars we can see huge
25$-$35 $\mu$m features (e.g., Hrivnak, Volk \& Kwok 2000) and very little
6$-$9, 9$-$13, and 15$-$20 $\mu$m features, contrary to what we observe in the
MCs. It seems that it is now well established that the carrier of unidentified
infrared emission (UIE) - both discrete and broad features - observed from
$\sim$3 to 20 $\mu$m in C-rich evolved stars should be carbon nanoparticles with
mixed aliphatic and aromatic structures similar to that of HAC dust (Kwok, Volk
\& Bernath 2001; Kwok \& Zhang 2011). Note that instead of HAC nanoparticles,
other organic matter with a complex mix of aliphatic and aromatic structures
(e.g., coal, petroleum fractions, quenched carbonaceous compounds, etc.) may be
present in the circumstellar envelopes of evolved stars and might be responsible
of the UIE (e.g., Kwok, Volk \& Bernath 2001). For example, coal and petroleum
fractions may explain a diverse set of aromatic and aliphatic features seen in
proto-PNe (e.g., Cataldo, Keheyan \& Heymann 2002). In this context, our {\it
Spitzer} spectra of fullerene PNe also show that a complex mixture of aliphatic
and aromatic species (e.g., HACs, PAH clusters, fullerenes, and small
dehydrogenated carbon clusters) is present in their circumstellar envelopes. 

Our interpretation is that the broad infrared features at 6$-$9, 9$-$13,
15$-$20, and 25$-$35 $\mu$m are intimately related with the formation of
fullerenes in PNe, being likely produced by a carbonaceous compound with a
mixture of aromatic and aliphatic structures (e.g., HAC) and/or their
decomposition products (e.g., fullerene precursors or intermediate products).
It is to be noted here that Kwok, Volk \& Bernath (2001) were the first to
suggest that the broad 6$-$9 and 9$-$13 $\mu$m features seen in proto-PNe may be
due to a carbonaceous compound with a mixed aromatic/aliphatic structure. The
6$-$9 and 15$-$20 $\mu$m features, which are almost exclusively seen in low
metallicity MC PNe (Figures 3 and 4), may be due to small dust grains (e.g., HAC
nanoparticles) while the 9$-$13 and 25$-$35 $\mu$m that are seen in all
metallicity environments (Figures 2 to 4) would be produced by big dust grains
(e.g., HAC aggregates). The broad 6$-$9 $\mu$m feature may be due to HACs, PAH
clusters or VSGs (see e.g., Tielens 2008) while the 15$-$20 $\mu$m emission may
be identified with the C-C-C bending modes of relatively large PAHs or PAH
clusters (Van Kerckhoven et al. 2000). On the other hand, and although the broad
9$-$13 and 25$-$35 $\mu$m features are usually attributed to SiC (the so-called
11.5 $\mu$m feature; e.g., Speck et al. 2009) and MgS (e.g., Hony et al. 2002),
respectively, these features may be due to bigger dust grains (e.g., HAC
aggregates). Garc\'{\i}a-Hern\'andez (2012) argued against the SiC and MgS
identification of these features in fullerene PNe and it will not be repeated
here. Basically, the 25$-$35 $\mu$m feature may be explained by HACs (see
Grishko et al. 2001; Garc\'{\i}a-Hern\'andez et al. 2010) and the spectral
characteristics (position, shape, see also Table 4) of the 9$-$13 $\mu$m
emission differ from those for the SiC 11.5$\mu$m feature. 

\subsection{Fullerene Emission vs. other IR Features}

In this Section we study the fullerene emission in our sample of fullerene PNe
versus other solid-state/molecular infrared features. The flux emitted in the
C$_{60}$ 18.9$\mu$m feature is roughly correlated with the fullerene content (or
C$_{60}$/C abundance values in Table 3; see also Figure 8) and can be taken as
representative of the fullerene emission. Almost all fullerene PNe in our sample
show some emission in the broad 9$-$13 $\mu$m feature\footnote{The only
exceptions are LMC 02 and LMC 99 with no significant 9$-$13 $\mu$m emission; see
Figure 3.} and in the 11.3$\mu$m PAH feature. However, not all PNe in our
sample display evident AIB emissions at $\sim$6.2 and 7.7 $\mu$m or in the very
broad 6$-$9 and 25$-$35 $\mu$m features, something that difficults the
exploration of any possible relation between the fullerene emission and the
carriers of the latter emission features. In particular, we prefer to exclude
(on a quantitative basis; see below) the very broad 6$-$9 and 25$-$35 $\mu$m
features from our study. This is because very few fullerene PNe display the
6$-$9 $\mu$m emission and the sensitivity of IRS drops dramatically for
$\lambda$ $\geq$36 $\mu$m and previous Infrared Space Observatory (ISO)
observations of PNe show that the 25$-$35 $\mu$m feature may go up to
wavelenghts longer than 40 $\mu$m (Hony et al. 2002). The central wavelengths
and integrated fluxes of the AIBs (usually attributed to PAHs) at
$\sim$6.2, 7.7, 11.3 $\mu$m as well as the broad 9$-$13 and 15$-$20 $\mu$m
features are given in Table 4.

Figure 8 displays the flux ratio C$_{60}$ 18.9$\mu$m/9$-$13$\mu$m versus the
fullerene abundances (in percentage), where we include the propagated error bars
corresponding to an individual flux error estimation of $\sim$15\%. Note that we
do not include the error bars of the fullerene abundances in order to avoid
dulling of Figure 8. The formal error in our estimation of the fullerene
abundances is difficult to calculate (e.g., different literature sources for the
C abundance, electron density, etc.) but it should be lower than 50\% in the
Magellanic Cloud sources - likely higher in the Galactic PNe (see \S 3.2).
Although most of the sources are concentrated in the lower left corner of the
diagram displayed in Figure 8, we find that the flux ratio C$_{60}$
18.9$\mu$m/9$-$13$\mu$m seems tentatively to increase with increasing
fullerene abundances. Also, two apparent outliers in Figure 8 seem to be PNe Tc
1 and SMC 15. The position of Tc 1 in this diagram, if real, may involve a
steeper correlation at the highest Galactic metallicities. However, the Tc 1's
{\it Spitzer} spectrum only covers the inner nebula while the {\it Spitzer}
apertures covers the full nebula in the rest of fullerene PNe displayed in
Figure 8. On the other hand, SMC 15 displays the highest electron density
(N$_{e}$=7500 cm$^{-3}$) among the PNe in the SMC (with an average N$_{e}$ of
$\sim$3500 cm$^{-3}$; see Table 2). It is to be noted here that by using the
average electron density for SMC 15 (3500 cm$^{-3}$), then a lower C$_{60}$/C
abundance of $\sim$0.03 \% is obtained. Thus, if we do not take into account Tc
1 and SMC 15, then Figure 8 would show an even tighter trend between the
C$_{60}$ 18.9$\mu$m/9$-$13$\mu$m flux ratio and fullerene abundances in
fullerene PNe. In summary, we conclude that there is a tentative observed
trend in Figure 8 but more detections of fullerenes in PNe are needed in order
to reach a definitive conclusion.

More interesting is that Figure 8 may be also interpreted as a consequence of
the destruction of the carrier of the 9$-$13 $\mu$m feature in those PNe
dominated by the fullerene emission. This would be expected if the carrier of
the broad 9$-$13 $\mu$m feature is a precursor of the fullerene
molecules\footnote{Note that the small amount of C$_{60}$ ($<$1\%, Table 3)
is with respect to the total gaseous carbon in the nebulae. We do not know what
fraction of carbon is locked in the carrier of the 9$-$13 $\mu$m feature. For
example, if the 9$-$13 $\mu$m feature's carrier is of solid-state origin (dust),
then only a small fraction of C (in comparison with gaseous carbon) is also
expected for the 9$-$13 $\mu$m feature's carrier.} and it is intimately related
with the fullerene formation process, as we have just interpreted in the
previous Section. As we have already mentioned in the Introduction, the strong
variability of the broad 9$-$13 $\mu$m emission from source to source (see Table
4) is more consistent with a HACs identification.

Figure 9 compares the fullerene emission with the AIB emission at 6.2 and 11.3
$\mu$m (e.g., the C$_{60}$ 18.9$\mu$m/AIB 11.3$\mu$m flux ratio versus the AIB
6.2$\mu$m/AIB 11.3$\mu$m flux ratio). We find two different groups in Figure 9.
A first group composed by three PNe (Tc 1, LMC 02, and SMC 24)\footnote{Note
that PN SMC 16 does not display the 6.2 and 7.7 $\mu$m AIBs and it would also
pertain to this first group of fullerene-dominated PNe.}, which are dominated by
the fullerene emission (C$_{60}$ 18.9$\mu$m/AIB 11.3$\mu$m $>$8) and show very
weak AIB emission at 6.2 and 11.3 $\mu$m (AIB 6.2$\mu$m/AIB 11.3$\mu$m $<$0.75).
These fullerene-dominated PNe are also characterized by the weakness of the
broad 9$-$13 $\mu$m feature compared to the fullerene emission. This suggests
that these three fullerene-dominated PNe may have a mix of aliphatic and
aromatic hydrocarbon materials of similar chemical composition (e.g.,. size and
degree of hydrogenation) and/or are under approximate physical conditions (e.g.., the
intensity of the UV radiation field) in their circumstellar envelopes. A second
group in Figure 9 is, however, formed by the rest of PNe in our sample,
displaying similar levels of fullerene emission but covering a wide range of AIB
6.2$\mu$m / AIB 11.3$\mu$m flux ratios, likely reflecting quickly changing
physical conditions (e.g., UV radiation field and/or the post-AGB shocks) and/or
chemical composition. The latter is consistent with the  relative strengths of
the IR features of amorphous organic nanoparticles with mixed aliphatic and
aromatic structures (e.g., HACs), which are known to be strongly dependent of
physical conditions and chemical composition (e.g., Scott et al. 1997; Grishko
et al. 2001). A similar result is obtained from Figure 10, where we display the
C$_{60}$ 18.9$\mu$m/AIB 11.3$\mu$m flux ratio versus the AIB 7.7$\mu$m/AIB
11.3$\mu$m flux ratio. Again, the same three fullerene-dominated PNe (although
Tc 1 does not show the 7.7 $\mu$m AIB) form a separated group in this latter
diagram, while the rest of objects show a wide range of AIB 7.7$\mu$m/AIB
11.3$\mu$m intensity ratios.

Finally, a few MC PNe (LMC 25, 48 in Figure 3 and SMC 13, 15, 18, and 20 in
Figure 4) show a broad 15$-$20 $\mu$m emission (see also Table 4). This broad
emission feature is usually attributed to the C-C-C bending modes of relatively
large PAHs or PAH clusters (e.g., Van Kerckhoven et al. 2000). The
15$-$20$\mu$m/11.3$\mu$m and 15$-$20$\mu$m/6.2$\mu$m flux ratios are displayed
in Figure 11 and they may be used to estimate the size (number of C atoms) of
the grains/PAH clusters responsible of this emission (see e.g., Van Kerckhoven
et al. 2000; Tappe et al. 2006, 2012 for more details). All members of the
15$-$20$\mu$m subgroup of MC PNe, with the exception of SMC 20, show AIB
11.3$\mu$m/15$-$20$\mu$m and AIB 6.2$\mu$m/15$-$20$\mu$m ratios lower than
$\sim$0.1 and 0.2, respectively, suggesting relatively large grains/PAHs
clusters of more than 4000 C atoms (see Tappe et al. 2006, 2012 and references
therein). Curiously, the observed characteristics of the 15$-$20 $\mu$m emission
in these sources - shape, integrated strength relative to the 6.2 and 11.3
$\mu$m AIBs, complete lack of the discrete emission features at 16.4, 17.0, and
17.8 $\mu$m - are very similar to the 15$-$20 $\mu$m bump observed in the
shock-induced dust processing environment of the supernova remnant N132D (Tappe
et al. 2006, 2012). Our interpretation is that these PAH clusters/amorphous
carbon nanoparticles survive the harsh conditions in the circumstellar envelope
long enough to be detected only in the low metallicity environments of MC PNe.
In our own Galaxy, however, the usual lack of PAH clusters-like 15$-$20$\mu$m
emission in fullerene PNe indicates that these PAH clusters/amorphous carbon
nanoparticles are quickly destroyed.

\subsection{Formation route of Fullerenes in PNe}

Some advances in our understanding of the fullerene formation process in PNe
have been made since the discovery of C$_{60}$ and C$_{70}$ fullerenes in the PN
Tc 1 by Cami et al. (2010). From the observational point of view, our recent
detections of fullerenes and graphene precursors (possible planar C$_{24}$) in
PNe have challenged our previous understanding on the formation of these complex
molecules, showing that they can form in H-rich circumstellar environments and
that they may coexist with PAH-like molecules (Garc\'{\i}a-Hern\'andez et al.
2010, 2011). The coexistente of a great diversity of aromatic and aliphatic
species in PNe with fullerenes supports the idea that fullerenes may form from
the photo-chemical processing (e.g., as a consequence of UV irradiation or
shocks) of small solid particles similar to that of HAC; the so-called ``HACs
fullerene formation scenario". The prevaling view was that the H-poor RCBs,
resembling the experimental conditions on Earth (Kroto et al. 1985), would be
the ideal astrophysical environments for fullerene formation. However, {\it
Spitzer} observations of RCB stars (Garc\'{\i}a-Hern\'andez, Rao \& Lambert
2011a,b) have shown that fullerene formation is inefficient in the highly
hydrogen-deficient circumstellar environment of most RCBs. Fullerenes are seen
in conjunction with strong PAH-like features in the two least H-deficient RCB
stars DY Cen and V854 Cen. In addition, the V854 Cen's IR spectrum evolved from
HACs to AIBs (e.g., from PAHs) and C$_{60}$ in a timescale of 10 years
only, supporting the ``HACs fullerene formation scenario". The fact that
previous {\it CLOUDY} photoionization models - by using BB and Rauch (2003)
stellar atmospheres as ionizing continuum - could not explain the low [Ne
III]/[Ne II] ratios observed in fullerene MC PNe with fast stellar winds (e.g.,
shocks) prompted the speculation that fullerenes may be formed by the
shock-induced decomposition of HAC material. This interpretation was consistent
with the low effective temperatures ($<$25,000 K) found in the other
astronomical environments such as the proto-PN IRAS 01005$+$7910, post-AGB
and RCB stars, and reflection nebulae, where fullerenes have been found so far
(see below). The UV radiation field in the latter astrophysical environments is
unlikely intense enough for the photo-chemical processing of HAC (see discussion
in Garc\'{\i}a-Hern\'andez, Rao \& Lambert 2011a). Indeed, the PAH clusters-like
15$-$20$\mu$m emission observed in fullerene PNe show a resemblance to that
observed in the shock-induced dust processing environment of the supernova
remnant N132D (Tappe et al. 2006, 2012). 

However, the new {\it CLOUDY} models presented in this paper (Section 3.1) show
that many of the observed [Ne III]/[Ne II] ratios may be explained by
photoionization when using model atmospheres with line and wind blanketing. This
result weakens the hypothesis of a shock-excited origin for the [Ne II] 12.8
$\mu$m emission. Although fast stellar winds (e.g., shocks) are generally
present in fullerene PNe\footnote{This is deduced from the common presence of
P-cygni profiles in their UV lines (see algo Garc\'{\i}a-Hern\'andez et al.
2011).}, the shock hypothesis cannot be directly tested with the actual data.
Furthermore, the winds of central stars of PNe are radiation-driven (Pauldrach
et al. 1986) and as such the wind velocity is expected to increase with the
central star temperature, allowing the development of wind-driven shock regions
when only the central star is sufficiently evolved (Villaver, Manchado \& 
Garc\'{\i}a-Segura 2002). Weaker shocks are also expected associated to the
propagation of the ionization front within the nebula. Figure 6 suggests that
the UV radiation from the central star can explain the low [Ne III]/[Ne II]
ratios observed in many fullerene PNe, putting some doubt to the suggestion of
the fullerenes being formed by the shock-induced HAC's decomposition. In
summary, this result together with the narrow range of T$_{eff}$
($\sim$30,000$-$45,000 K) displayed by our sample sources in Figure 6 indicate
common fullerene formation conditions in PNe and suggest that perhaps the UV
irradiation is photo-chemically processing the small solid particles with mixed
aromatic and aliphatic structures similar to that of HAC dust and that may form
the fullerenes. This interpretation is consistent with the early suggestion
by Kwok, Volk \& Bernath (2001), who emphasizes the important role of
photo-chemistry in the evolution from proto-PNe to PNe. 

From the experimental point of view, we would like to emphasize the recent
finding by Duley \& Hu (2012) that HAC nanoparticles prepared under a variety of
conditions can show IR features coincident with the four C$_{60}$ IR features,
and that may be attributed to fullerene precursors or proto-fullerenes. The
presence of proto-fullerenes takes place in conjunction with the 16.4 $\mu$m
feature. The 16.4 $\mu$m feature is usually present in the proto-PN IRAS
01005$+$7910, post-AGB and RCB stars, and reflection nebulae (all of them with
T$_{eff}$$<$30,000 K), likely indicating the presence of proto-fullerenes
(Garc\'{\i}a-Hern\'andez et al. 2012) as a previous stage in the evolution and
dehydrogenation of UV irradiated HAC dust in fullerene PNe. More recently, Dunk
et al. (2012) propose a Closed Network Growth (CNG) mechanism of fullerenes,
where larger fullerenes may growth from preexisting C$_{60}$ molecules and this
formation route can work under the presence of hydrogen. CNG of fullerenes would
be consistent with the ``HAC's fullerene formation scenario" and it may work in
the complex circumstellar environments of fullerene PNe. In this context, the
smaller C$_{50}$, C$_{60}$, and C$_{70}$ would be supplied by the photo-chemical
processing of HAC dust.

\section{Conclusions}
In this paper we present a detailed study of the {\it Spitzer} infrared spectra
of sixteen fullerene PNe. We report the first detection of fullerenes
in the Galactic PN M 1-60 as well as the presence of the unusual set of infrared
features at $\sim$6.6, 9.8, and 20 $\mu$m (attributed to possible planar
C$_{24}$) in the PN K 3-54. 

The observed C$_{60}$ intensity ratios in the Galactic PNe with fullerene
emission confirm our previous finding in the Magellanic Clouds that fullerenes -
probably in solid-state - in PNe are not excited by the UV radiation from the
central star. In addition, new {\it CLOUDY} photoionization models show that
line- and wind-blanketed model atmospheres can explain many of the observed
[NeIII]/[NeII] ratios by photoionization and that PNe containing fullerenes
display a rather narrow range in effective temperature ($\sim$30,000$-$45,000
K). This shows that fullerene PNe share similar physical conditions for the
fullerene formation and that perhaps is the UV radiation from the central star
that is processing the circumstellar dust grains - e.g., small solid particles
containing aromatic, aliphatic, fullerene, and grahene structures similar to
that of HAC dust - and forming fullerenes.

We have studied the fullerene emission with respect to other infrared emission
features such as the classical AIBs (usually attributed to PAHs) or the broad
emission features at 6$-$9, 9$-$13, 15$-$20, and 25$-$35 $\mu$m that we interpret
as produced by a carbonaceous compound with a mixture of aromatic and aliphatic
structures (e.g., HAC, coal, petroleum fractions, etc.) and/or their
decomposition products (e.g., fullerene precursors or intermediate products).
Thus, fullerenes and proto-graphene in PNe may be formed from the photo-chemical
processing of such a carbonaceous compound, which should be very abundant in
their circumstellar envelopes. Our interpretation is consistent with the
early seminal work by Kwok, Volk \& Bernath (2001). 

The PNe targets that show fullerene emission belong to very different
metallicity environments such as our own Galaxy and the Magellanic Clouds.  The
detection rate of fullerenes in carbon-rich PNe is found to increase with
decreasing metallicity, something that seems to be strongly linked to the
limited dust processing in Magellanic Cloud PNe when compared to the Galactic
PNe. It is important, however, to acknowledge the fact that all the original
samples of PNe from which the fullerene detections come from have their own
selection biases and cannot be considered complete by any means. The fact that
in general the {\it Spitzer} spectra of Magellanic Cloud PNe are usually
dominated by small dust grains (e.g., HAC and/or PAH clusters or even HAC
decomposition products), while their Galactic counterparts display dust emission
mainly from big grains (e.g.,. HAC aggregates) also points towards a higher
probability of fullerene detection with decreasing metallicity under our
proposed mechanism for fullerene formation. Our estimated fullerene abundances
and temperatures are similar in the various metallicity environments analyzed.
When possible, we have argued that fullerene PNe likely evolve from
low-mass progenitors (main sequence masses $<$ 1.5 M$_{\sun}$) and are usually
low-excitation objects slowly evolving towards the white dwarf stage. 



\acknowledgments

We acknowledge support from the Faculty of the European Space Astronomy Centre
(ESAC). D.A.G.H.and A.M. acknowledge support for this work provided by the
Spanish Ministry of Economy and Competitiveness under grant AYA$-$2011$-$27754.
D.A.G.H. also thanks his son Mateo for his great patience during the final
writing of the paper and C. Morisset for pointing out that the [NeIII]/[NeII]
ratios might be explained with models including line and wind blanketing. EV
acknowledges the support provided  by the Spanish Ministry of Science and
Innovation (MICINN)  under grant AYA$-$2010$-$20630 and to the Marie Curie
FP7$-$  People$-$RG268111. L.S. and R.A.S. acknowledge support by NASA through
awards for programs GO 20443 and 50261 issued by JPL$/$Caltech. Thanks to James
Davies for his help with the data analysis and to Susana Iglesias-Groth for the
construction of the excitation diagrams. This work is based on observations 
made with the Spitzer Space Telescope, which is operated by the Jet Propulsion
Laboratory, California Institute of Technology, under NASA contract 1407.



{\it Facilities:} \facility{Spitzer:IRS}.

\clearpage

\begin{deluxetable}{ccccccccc}
\tabletypesize{\scriptsize}
\tablecaption{Mid-IR C$_{60}$, C$_{70}$, and planar C$_{24}$ features in fullerene PNe$^{a}$.\label{tbl-1}}
\tablewidth{0pt}
\tablehead{
\colhead{} & \colhead{} & \colhead{} & \colhead{} & \colhead{C$_{60}$} & \colhead{} &
\colhead{} & \colhead{} & \colhead{}\\
\hline
\colhead{Object} & \colhead{$\lambda$} & \colhead{Flux} &
\colhead{$\lambda$} & \colhead{Flux} &
\colhead{$\lambda$} & \colhead{Flux} &
\colhead{$\lambda$} & \colhead{Flux} \\
\colhead{} & \colhead{($\mu$m)} & \colhead{(Wcm$^{-2}$)} &
\colhead{($\mu$m)} & \colhead{(Wcm$^{-2}$)}&
\colhead{($\mu$m)} & \colhead{(Wcm$^{-2}$)} &
\colhead{($\mu$m)} & \colhead{(Wcm$^{-2}$)} 
}
\startdata
SMC 13 & 7.13& 4.00e-21($\pm$1.00) & 8.66& 2.53e-21($\pm$0.63) & 17.50& 7.59e-22($\pm$1.52) & 18.98& 3.66e-21($\pm$0.55) \\
SMC 15 & 7.09& 4.01e-21($\pm$1.20) & 8.43& 1.49e-21($\pm$0.45) & 17.49& 9.07e-22($\pm$1.81) & 18.98& 3.03e-21($\pm$0.45) \\
SMC 16 & 7.02& 6.64e-21($\pm$1.33) & 8.57& 2.71e-21($\pm$0.81) & 17.40& 3.78e-21($\pm$0.57) & 18.97& 8.33e-21($\pm$1.25) \\
SMC 18 & 7.03& 5.46e-21($\pm$1.37) & 8.68& 2.66e-21($\pm$0.57) & 17.41& 2.25e-21($\pm$0.34) & 18.96& 5.16e-21($\pm$0.77) \\
SMC 20 & 7.00& 1.57e-21($\pm$0.47) & 8.40& 4.18e-22($\pm$1.25) &$\dots$& $\dots$	    & 19.01& 3.27e-21($\pm$0.65) \\
SMC 24 & 7.04& 6.22e-21($\pm$1.24) & 8.62& 4.07e-21($\pm$1.22) & 17.52& 4.19e-21($\pm$0.63) & 18.94& 7.93e-21($\pm$1.19) \\
SMC 27 & 7.27& 2.65e-21($\pm$1.06) & 8.57& 1.09e-21($\pm$0.44) & 17.34& 1.14e-21($\pm$0.34) & 18.99& 2.29e-21($\pm$0.46) \\
LMC 02 & 7.04& 7.43e-21($\pm$1.49) & 8.58& 3.27e-21($\pm$0.98) & 17.44& 3.31e-21($\pm$0.50) & 18.92& 4.66e-21($\pm$0.70) \\
LMC 25 & 6.92& 1.43e-21($\pm$0.36) & 8.51& 2.26e-21($\pm$0.57) & 17.39& 6.05e-21($\pm$1.21) & 18.97& 8.77e-21($\pm$1.32) \\
LMC 48 & 7.03& 4.47e-21($\pm$1.12) & 8.65& 2.85e-21($\pm$0.71) & 17.45& 1.48e-20($\pm$0.30) & 18.97& 1.31e-20($\pm$0.20) \\
LMC 99 & 6.96& 3.13e-21($\pm$0.94) & 8.70& 5.05e-21($\pm$1.01) & 17.36& 3.81e-21($\pm$0.57) & 19.00& 5.27e-21($\pm$0.79) \\
Tc 1   & 7.02& 1.68e-18($\pm$0.25) & 8.51& 5.96e-19($\pm$0.89) & 17.33& 8.78e-19($\pm$1.32) & 18.89& 1.65e-18($\pm$0.25) \\
M 1-20 & 7.02& 1.04e-19($\pm$0.21) & 8.58& 6.79e-20($\pm$1.36) & 17.35& 6.70e-20($\pm$1.01) & 18.96& 1.36e-19($\pm$0.20) \\
M 1-12 & 6.99& 2.24e-19($\pm$0.34) & 8.58& 1.07e-19($\pm$0.16) & 17.37& 6.48e-20($\pm$0.97) & 18.94& 1.29e-19($\pm$0.19) \\
M 1-60 & 7.00& 9.59e-20($\pm$1.92) & 8.61& 7.13e-20($\pm$1.43) & 17.34& 3.73e-19($\pm$0.56) & 19.15& 6.82e-19($\pm$1.02) \\
K 3-54 & 7.01& 5.79e-20($\pm$0.87) & 8.48& 3.83e-21($\pm$0.77) & 17.38& 2.58e-20($\pm$0.39) & 18.86& 4.06e-20($\pm$0.61) \\
\hline
 &  &  &  & C$_{70}$ &  &  &  & \\
\hline
Object& $\lambda$ & Flux & $\lambda$ & Flux & $\lambda$ & Flux & $\lambda$ & Flux\\
 & ($\mu$m) & (Wcm$^{-2}$) & ($\mu$m) & (Wcm$^{-2}$)& ($\mu$m) & (Wcm$^{-2}$) & ($\mu$m) & (Wcm$^{-2}$) \\
\hline
SMC 24 & 14.93 & 2.45e-21($\pm$0.73) & 15.56 & 7.49e-21($\pm$1.50) & 17.77	& 2.96e-21($\pm$0.59)&$\dots$  & $\dots$\\
LMC 02 &$\dots$&  $\dots$		 & 15.57 & 6.37e-22($\pm$1.27) & 17.72	& 1.56e-22($\pm$0.47)& 18.73   & 8.02e-22($\pm$1.60)\\
Tc 1   & 12.53 & 4.24e-20($\pm$0.85) & 14.79 & 8.34e-20($\pm$1.67) & 15.55      & 1.01e-19($\pm$0.20)& 21.75   & 2.71e-20($\pm$0.68) \\ 
\hline
 &  &  &  & C$_{24}$ &  &  &  & \\
\hline
Object& $\lambda$ & Flux & $\lambda$ & Flux & $\lambda$ & Flux & $\lambda$ & Flux\\
 & ($\mu$m) & (Wcm$^{-2}$) & ($\mu$m) & (Wcm$^{-2}$)& ($\mu$m) & (Wcm$^{-2}$) & ($\mu$m) & (Wcm$^{-2}$) \\
\hline
SMC 24 & 6.62    &  7.25e-21($\pm$1.45)  & 9.79 & 2.20e-21($\pm$0.44) & 20.02 & 5.03e-21($\pm$1.01)  &$\dots$  & $\dots$\\
LMC 02 & $\dots$ &  $\dots$                  & 9.79 & 2.66e-21($\pm$0.53) & 20.08 & 9.52e-22($\pm$1.90)  &$\dots$  & $\dots$\\
K 3-54 & 6.59    &  1.19e-20($\pm$0.18)  & 9.60 & 5.56e-21($\pm$1.12) & 20.02 & 1.14e-20($\pm$0.17)  &$\dots$  & $\dots$\\
\enddata
\tablenotetext{a}{The first eleven PNe pertain to the Magellanic Clouds while
the other five objects are located in our own Galaxy. Estimated flux errors
(between brackets) are always less than $\sim$30-40\%. The estimated flux errors
are lower (typically $\sim$15\%) for the Galactic PNe as well as for the
strongest fullerene transitions (e.g., at 17.4 and 18.9 $\mu$m) in PNe of the
Magellanic Clouds (see text for more details).}
\end{deluxetable}

\clearpage

\begin{deluxetable}{ccccccccc}
\tabletypesize{\scriptsize}
\tablecaption{The sample of fullerene PNe.\label{tbl-2}}
\tablewidth{0pt}
\tablehead{
\colhead{Object} &  \colhead{T$_{eff}$} &  \colhead{log(F(H$_{\beta}$))} & \colhead{A(C)} 

&  \colhead{T$_{e}$} & \colhead{N$_{e}$} & \colhead{Ref.$^{a}$} &  \colhead{M$_{H}$} & \colhead{M$_{C}$}   \\
\colhead{} &  \colhead{(K)}  & \colhead{} & \colhead{log(C/H)+12} &
\colhead{(10$^{4}$ K)} &  \colhead{(cm$^{-3}$)} 
& \colhead{} & \colhead{(M$_{\odot}$)} & \colhead{(10$^{-3}$ M$_{\odot}$)}
}
\startdata
SMC 13       & 31,300 &  -12.59 &  8.73 &  1.28  & 2900    &  (1,2,3,4,4)        &  0.702  & 4.520 \\
SMC 15       & $\dots$&  -12.45 &  8.26 &  1.41  & 7500    &  ($\dots$,5,3,6,7)	 &  0.270  & 0.588 \\
SMC 16       & $\dots$&  -12.74 &  8.19 &  1.18  & 4400    &  ($\dots$,5,3,6,7)	 &  0.210  & 0.392 \\
SMC 18       & $\dots$&  -12.66 &  8.31 &  1.18  & 3600    &  ($\dots$,2,3,4,4)	 &  0.381  & 0.934 \\
SMC 20       & $\dots$&  -12.47 &  8.25 &  1.38  & 3900    &  ($\dots$,2,3,4,4)	 &  0.497  & 1.061 \\
SMC 24$^{c}$ & 37,800 &  -12.66 &  8.18 &  1.16  & 2800    &  (1,2,3,4,4)	 &  0.411  & 0.747 \\
SMC 27       & 43,300 &  -12.51 &$\dots$&  1.27  & 3650    &  (1,2,$\dots$,4,4)	 &  0.471  & $\dots$ \\
LMC 02$^{c}$ & 39,000 &  -13.18 &  8.14 &  1.21  & 5000    &  (8,7,9,6,8)	 &  0.048  & 0.079 \\
LMC 25       & 33,700 &  -12.39 &  8.29 &  1.56  & 3300    &  (10,11,12,13,13)	 &  0.626  & 1.465 \\ 
LMC 48       & $\dots$&  -12.48 &  8.40 &  1.00$^{b}$ & 1900 & ($\dots$,5,12,$\dots$,6)  &  0.844  & 2.540 \\
LMC 99       & $\dots$&  -12.54 &  8.77 &  1.22  & 3600    & ($\dots$,14,15,6,6) &  0.434  & 3.040 \\
Tc 1$^{c}$   & 34,100 &  -10.73 &  8.67 & 0.90         & 3000       &  (16,17,18,18,18)                &  0.0535      & 0.300  \\
M 1-20$^{d}$ & 53,000 &  -11.93 &  8.55 & 1.11         & 7500       &  (19,17,$\dots$,20,21)           &  0.1558      & 0.715  \\
M 1-12$^{d}$ & 29,100 &  -11.60 &  8.82 & 0.95         & 7400       &  (16,17,$\dots$,22,23)           &  0.1708      & 1.354  \\
M 1-60$^{d}$ & 74,400 &  -12.28 &  8.82 & 0.92$^{e}$   & 9200$^{e}$ &  (16,17,$\dots$,24,24)           &  0.1752      & 1.384  \\
K 3-54$^{d}$ & $\dots$&  -13.30 &  8.55 & 1.02         & 5000$^{f}$ &  ($\dots$,25,$\dots$,17,$\dots$) &  0.0619      & 0.263  \\
\enddata
\tablenotetext{a}{References for T$_{eff}$, log(F(H$_{\beta}$)), C abundance, T$_{e}$, and N$_{e}$. Note that literature T$_{eff}$ values are mostly HI Zanstra temperatures (see Section 3.1).}
\tablenotetext{b}{T$_{e}$ for LMC 48 is assumed to be 10$^{4}$ K.}
\tablenotetext{c}{High-resolution Spitzer/IRS spectra were obtained and it was possible to detect C$_{70}$ unambiguously.}
\tablenotetext{d}{The C abundance for these PNe is assumed from the average values of the height above the Galactic plane (see text).}
\tablenotetext{e}{T$_{e}$ and N$_{e}$ for M 1-60 are derived by using the line fluxes given by Girard et al. (2007).}
\tablenotetext{f}{N$_{e}$ for K 3-54 is assumed to be 5000 cm$^{-3}$.}
\tablerefs{(1) Villaver et al. (2004); (2) Stanghellini et al. (2003); (3) Stanghellini et al. (2009); (4) Shaw et al.
(2010); (5) Shaw et al. (2006); (6) Leisy \& Dennefeld (2006); (7) Dopita et al. (1988); (8) Herald \& Bianchi (2004);
(9) Milanova \& Kholtygin (2009); (10) Villaver et al. 2003; (11). Stanghellini et al. (2002); (12) Stanghellini et al.
(2005); (13) Shaw et al. (2013, in prep.); (14) R. Shaw (2011, priv. comm.); (15) Leisy \& Dennefeld (1996); (16)
Preite-Mart\'{\i}nez et al.(1989); (17) Cahn et al. (1992); (18) Pottasch et al. (2011); (19) Kaler \& Jacoby (1991);
(20) Chiappini et al.(2009); (21) Richer et al. (2008); (22) Henry et al. (2010); (23) Stanghellini \& Kaler (1989); (24) Girard et al.
(2007); (25) Acker et al. (1991).}
\end{deluxetable}

 \clearpage

\begin{deluxetable}{cccccccccccc}
\tabletypesize{\scriptsize}
\setlength{\tabcolsep}{0.02in} 
\tablecaption{C$_{60}$ intensity ratios and fullerene abundances and temperatures.\label{tbl-2}}
\tablewidth{0pt}
\tablehead{ \colhead{Object} &  
\colhead{F(7.0)/F(18.9)$^{a}$} &
\colhead{F(8.5)/F(18.9)$^{a}$} &
\colhead{F(17.4)/F(18.9)$^{a}$} & \colhead{[NeIII]/[NeII]} &
\colhead{C$_{60}$/C} &  \colhead{f$_{C60}$$^{b}$} & 
\colhead{T$_{C60}$} & \colhead{C$_{70}$/C}&  \colhead{f$_{C70}$$^{b}$} & \colhead{T$_{C70}$} & \colhead{C$_{70}$/C$_{60}$}\\

\colhead{} & \colhead{} &\colhead{} &\colhead{} &\colhead{} &

\colhead{(\%)}& \colhead{(\%)} & \colhead{(K)} & \colhead{(\%)} & \colhead{(\%)}& \colhead{(K)} & \colhead{}
}
\startdata
SMC 13  &   0.16  &0.32&    0.21    & 2.89  & 0.003  & 15$\pm$10 & 453$\pm$50 &$\dots$	&$\dots$ & $\dots$ & $\dots$	 \\
SMC 15  &   0.20  &0.49&    0.30    & 0.49  & 0.07   & 15$\pm$10 & 290$\pm$20 &$\dots$	&$\dots$ & $\dots$ & $\dots$	 \\
SMC 16  &   0.24  &0.33&    0.45    & 0.21  & 0.12   & 30$\pm$10 & 414$\pm$50 &$\dots$	&$\dots$ & $\dots$ & $\dots$     \\
SMC 18  &   0.16  &0.52&    0.44    & 1.16  & 0.03   & 15$\pm$10 & 365$\pm$35 &$\dots$	&$\dots$ & $\dots$ & $\dots$	 \\
SMC 20  &   0.05  &0.13& $\dots$    & 0.41  & 0.03   & 10$\pm$10 & 228$\pm$30 &$\dots$	&$\dots$ & $\dots$ & $\dots$	 \\
SMC 24  &   0.47  &0.18&    0.53    & 3.33  & 0.07   & 60$\pm$20 & 514$\pm$30 & 0.001	& 20$\pm$10& 383$\pm$60 &  0.02       \\
SMC 27  &   0.58  &0.48&    0.50    & 9.69  &$\dots$ & 50$\pm$10 & 553$\pm$40 &$\dots$	&$\dots$ & $\dots$ & $\dots$	 \\
LMC 02  &   0.48  &0.37&    0.71    & 0.42  & 0.29   & 30$\pm$10 & 502$\pm$20 & 0.07	& 10$\pm$10& 324$\pm$10 &  0.21		 \\	 
LMC 25  &   0.08  &0.26&    0.69    &13.19  & 0.02   & 50$\pm$20 & 300$\pm$25 &$\dots$	&$\dots$ & $\dots$ & $\dots$	 \\
LMC 48  &   0.07  &0.22&    1.13    &11.46  & 0.01   & 20$\pm$10 & 270$\pm$20 &$\dots$	&$\dots$ & $\dots$ & $\dots$	 \\
LMC 99  &   0.30  &0.56&    0.72    &16.82  & 0.02   & 50$\pm$10 & 446$\pm$25 &$\dots$	&$\dots$ & $\dots$ & $\dots$	 \\
Tc 1    &   0.26  &0.36&    0.53    & 0.05  & 0.04   & 25$\pm$10 & 415$\pm$30 & 0.005   & 40$\pm$10& 314$\pm$30 & 0.10	 \\
M 1-20  &   0.38  &0.50&    0.49    &12.22  & 0.02   & 50$\pm$10 & 480$\pm$50&$\dots$ &$\dots$ & $\dots$ & $\dots$     \\ 
M 1-12  &   0.78  &0.58&    0.50    &$\dots$& 0.02   & 45$\pm$15 & 660$\pm$100&$\dots$        &$\dots$ & $\dots$ & $\dots$     \\
M 1-60  &   0.04  &0.11&    0.55    &12.94  & 0.05   & 30$\pm$10 & 260$\pm$30 &$\dots$        &$\dots$ & $\dots$ & $\dots$     \\
K 3-54  &   0.29  &0.25&    0.64    & 0.02  & 0.01   & 20$\pm$10 & 360$\pm$50 &$\dots$        &$\dots$ & $\dots$ & $\dots$     \\
\enddata
\tablenotetext{a}{The ratio of fluxes in the C$_{60}$ bands with the wavelenghts
in brackets. Note that we list the C$_{60}$ intensity ratios obtained by
correcting the observed emission fluxes at 7 $\mu$m (also at 8.5 $\mu$m for some
PNe in our sample) by the C$_{60}$ contribution factors (f$_{C60}$) derived from
the Boltzmann excitation diagrams (see text for more details).}
\tablenotetext{b}{C$_{60}$ and C$_{70}$ contributions to the observed 7$\mu$m flux.}
\end{deluxetable}

\clearpage

\begin{landscape}

\begin{deluxetable}{ccccccccccc}
\tabletypesize{\scriptsize}
\tablecaption{Broad 9$-$13 and 15$-$20 $\mu$m emission and 6.2, 7.7, and 11.3 $\mu$m AIBs in fullerene PNe$^{a}$.\label{tbl-1}}
\tablewidth{0pt}
\tablehead{
\colhead{Object} & \colhead{$\lambda$} & \colhead{Flux} & \colhead{$\lambda$} & \colhead{Flux} &
\colhead{$\lambda$} & \colhead{Flux} &
\colhead{$\lambda$} & \colhead{Flux} &
\colhead{$\lambda$} & \colhead{Flux} \\
\colhead{} & \colhead{($\mu$m)} & \colhead{(Wcm$^{-2}$)} & \colhead{($\mu$m)} & \colhead{(Wcm$^{-2}$)} &
\colhead{($\mu$m)} & \colhead{(Wcm$^{-2}$)}&
\colhead{($\mu$m)} & \colhead{(Wcm$^{-2}$)} &
\colhead{($\mu$m)} & \colhead{(Wcm$^{-2}$)} 
}
\startdata
SMC 13 & 11.35   & 4.78e-20($\pm$0.72)& 19.06   & 1.87e-20($\pm$0.28) & 6.25  & 3.74e-21($\pm$0.94) & 7.80    & 8.47e-21($\pm$2.12) & 11.23 & 4.86e-21($\pm$1.22) \\
SMC 15 & 11.24   & 2.51e-19($\pm$0.38)& 18.29   & 1.01e-19($\pm$0.15) & 6.21  & 2.92e-21($\pm$0.88) & 7.71    & 1.17e-20($\pm$0.29) & 11.34 & 3.52e-21($\pm$0.88)\\
SMC 16 & 11.56   & 2.50e-20($\pm$0.38)& $\dots$ & $\dots$             &$\dots$& $\dots$	    & $\dots$ & $\dots$             & 11.31 & 6.75e-23($\pm$2.70)\\
SMC 18 & 11.20   & 1.63e-19($\pm$0.24)& 18.53   & 1.56e-19($\pm$0.23) &$\dots$& $\dots$	    & 7.74    & 4.42e-21($\pm$1.33) & 11.26 & 3.85e-21($\pm$1.15)\\
SMC 20 & 10.94   & 2.68e-19($\pm$0.40)& 18.05   & 3.71e-20($\pm$0.56) & 6.27  & 4.21e-21($\pm$1.05) & 7.92    & 4.72e-21($\pm$1.42) & 11.33 & 3.42e-21($\pm$1.03)\\
SMC 24 & 11.34   & 3.85e-20($\pm$0.58)& $\dots$ & $\dots$             & 6.21  & 5.36e-22($\pm$2.14) & 7.71    & 3.67e-21($\pm$1.65) & 11.30 & 9.24e-22($\pm$2.31)\\
SMC 27 & 11.56   & 3.09e-20($\pm$0.46)& $\dots$ & $\dots$             &$\dots$& $\dots$	            & $\dots$ & $\dots$             & 11.28 & 2.03e-21($\pm$0.41)\\
LMC 02 & $\dots$ & $\dots$                & $\dots$ & $\dots$         & 6.17  & 1.71e-22($\pm$0.69) & 7.67    & 4.55e-22($\pm$1.59) & 11.29 & 3.86e-22($\pm$1.55)\\
LMC 25 & 11.38   & 1.74e-19($\pm$0.26)& 18.36   & 1.20e-19($\pm$0.18) & 6.28  & 1.73e-20($\pm$0.26) & 7.67    & 2.95e-20($\pm$0.59) & 11.28 & 4.23e-21($\pm$0.85)\\
LMC 48 & 11.41   & 1.37e-19($\pm$0.21)& 17.96   & 1.25e-19($\pm$0.19) & 6.25  & 9.87e-21($\pm$1.48) & 7.55    & 1.90e-20($\pm$0.38) & 11.31 & 3.26e-21($\pm$0.65)\\
LMC 99 & $\dots$ & $\dots$                & $\dots$ & $\dots$         & 6.27  & 1.46e-20($\pm$0.22) & 7.87    & 2.68e-20($\pm$0.40) & 11.30 & 1.44e-20($\pm$0.22)\\
Tc 1   & 11.75   & 2.92e-18($\pm$0.44)& $\dots$ & $\dots$             & 6.22  & 6.09e-20($\pm$1.22) & $\dots$ & $\dots$		    & 11.28 & 1.05e-19($\pm$0.16) \\
M 1-20 & 11.69   & 5.20e-18($\pm$0.78)& $\dots$ & $\dots$             & 6.25  & 2.01e-19($\pm$0.30) & 7.58    & 2.63e-19($\pm$0.39) & 11.29 & 8.93e-20($\pm$1.34) \\
M 1-12 & 11.46   & 6.53e-18($\pm$0.98)& $\dots$ & $\dots$             & 6.23  & 2.32e-19($\pm$0.35) & 7.57    & 4.49e-19($\pm$0.67) & 11.26 & 8.83e-20($\pm$1.32) \\
M 1-60 & 11.28   & 2.91e-18($\pm$0.44)& $\dots$ & $\dots$             & 6.26  & 3.08e-19($\pm$0.46) & 7.83    & 1.89e-19($\pm$0.28) & 11.30 & 3.37e-19($\pm$0.51) \\
K 3-54 & 11.62   & 6.94e-19($\pm$1.04)& $\dots$ & $\dots$             & 6.27  & 3.46e-20($\pm$0.69) & 7.58    & 5.95e-20($\pm$0.89) & 11.29 & 1.24e-20($\pm$0.19) \\
\enddata
\tablenotetext{a}{Estimated flux errors (between brackets) are always less than
$\sim$30-40\%. The estimated flux errors are lower (typically $\sim$15\%) for
the Galactic PNe as well as for the broad features in Magellanic Cloud PNe.}
\end{deluxetable}

\end{landscape}

\clearpage

\begin{figure}
\includegraphics[angle=90,scale=.33]{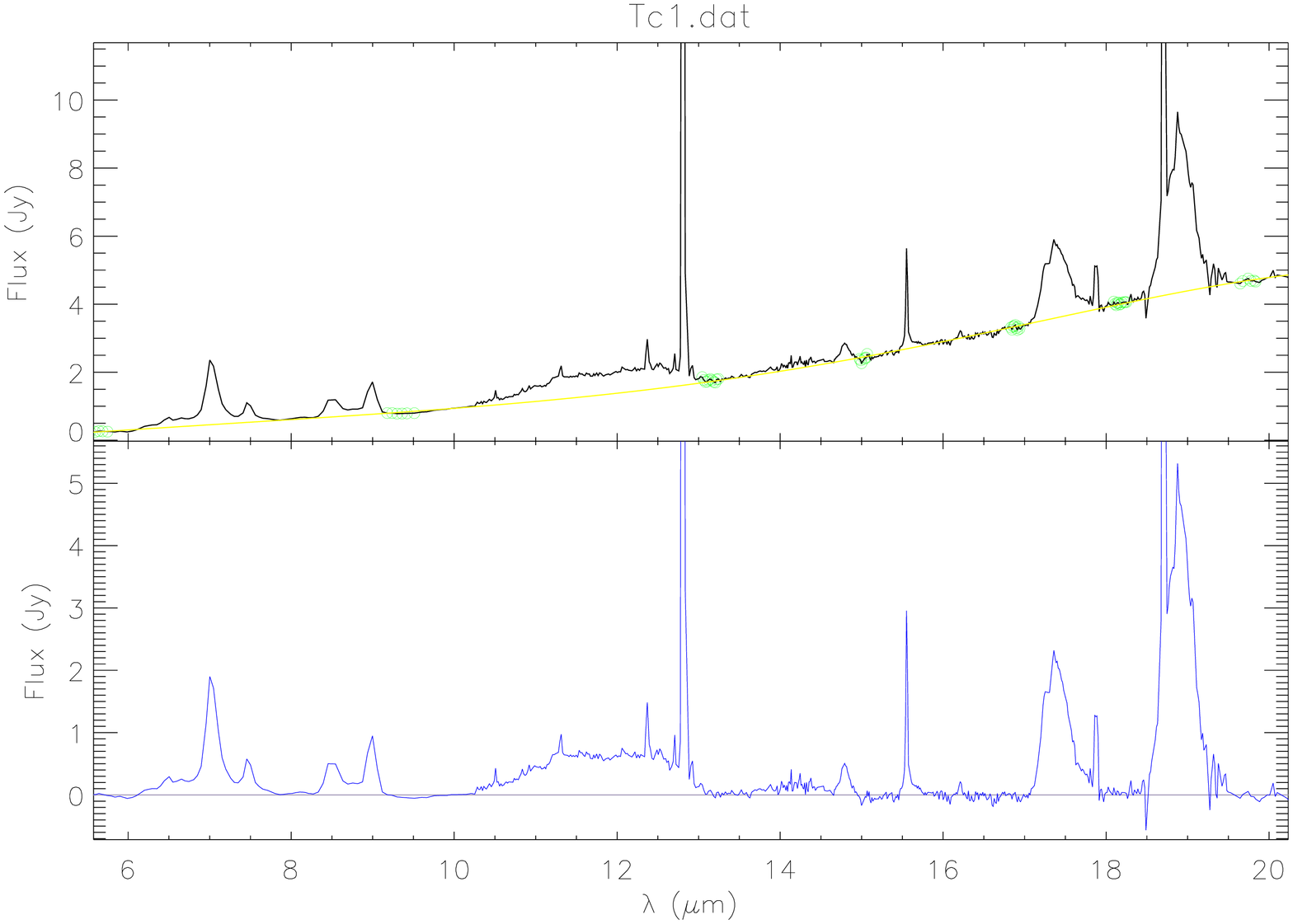}%
\includegraphics[angle=90,scale=.33]{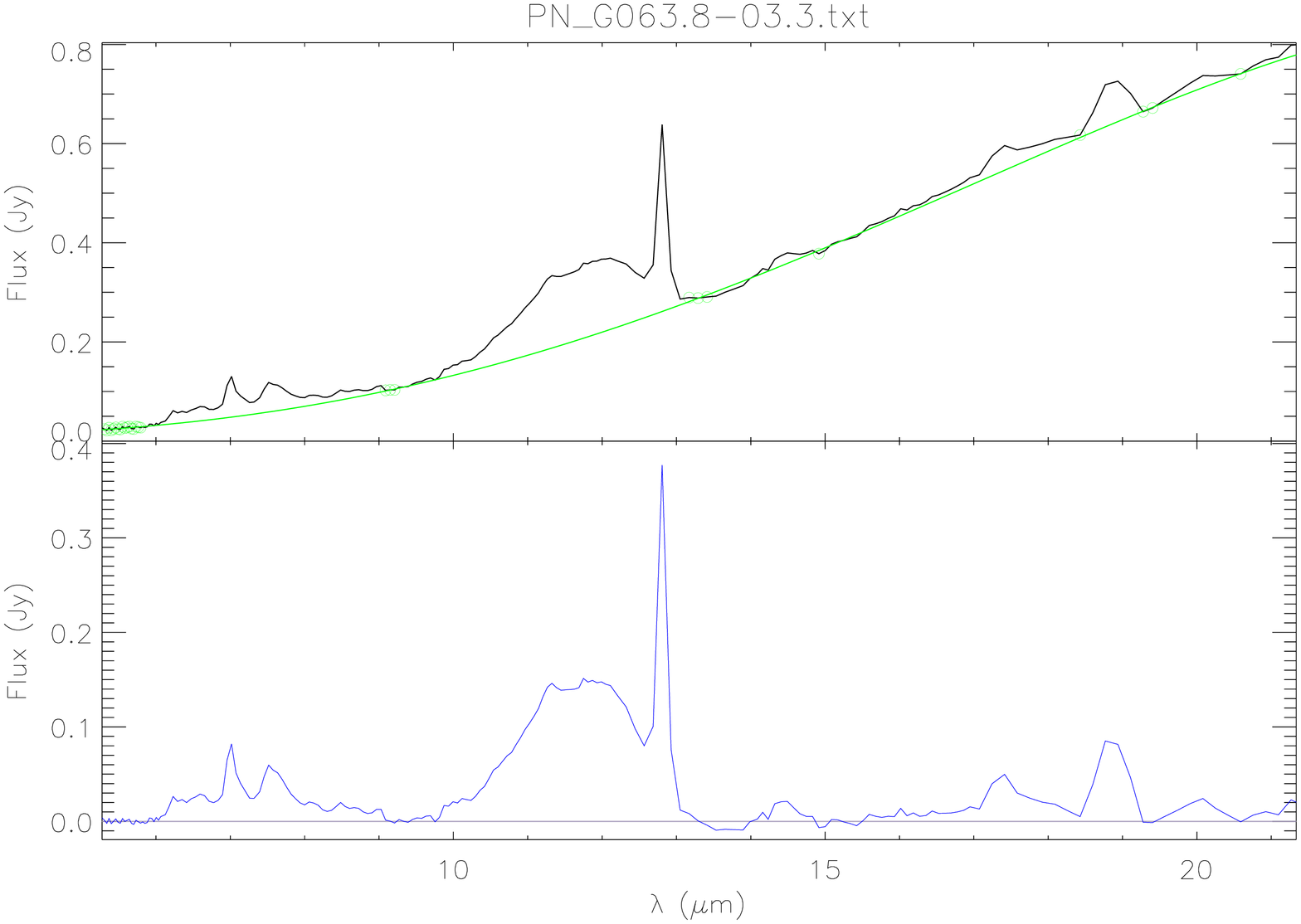}
\caption{Illustrative examples of the polynomical fits made to the dust
continuum and the corresponding residual spectra for the fullerene
PNe Tc 1 (left panel) and K 3-54 (right panel). Both top panels show the
observed Spitzer/IRS spectra (in black) together with a polynomial fit to
continuum points free grom any gas and dust feature. The corresponding residual
or dust continuum subtracted spectra (in blue) are shown in the bottom panels.
\label{fig9}}
\end{figure}

\begin{figure}
\includegraphics[angle=0,scale=.70]{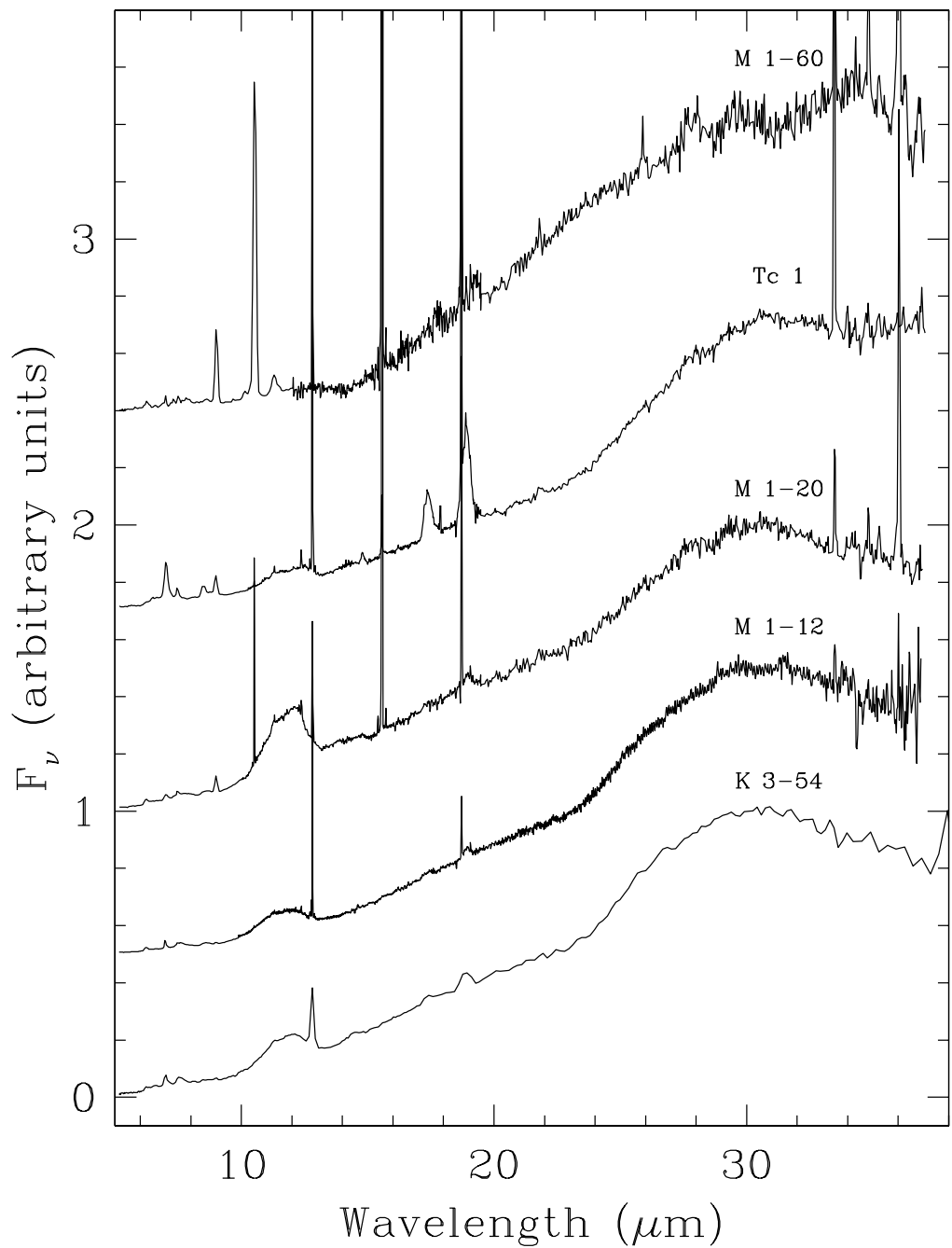}
\caption{Spitzer/IRS spectra in the wavelength $\sim$5$-$38 $\mu$m for the five
Galactic fullerene PNe in our sample. The spectra are normalized at
30 $\mu$m and the vertical axis has been artificially displaced for clarity. Note that
the strongest C$_{60}$ features at 17.4 and 18.9 $\mu$m are clearly detected
superimposed on the dust continuum thermal emission.
\label{fig2}}
\end{figure}

\clearpage

\begin{figure}
\includegraphics[angle=0,scale=.70]{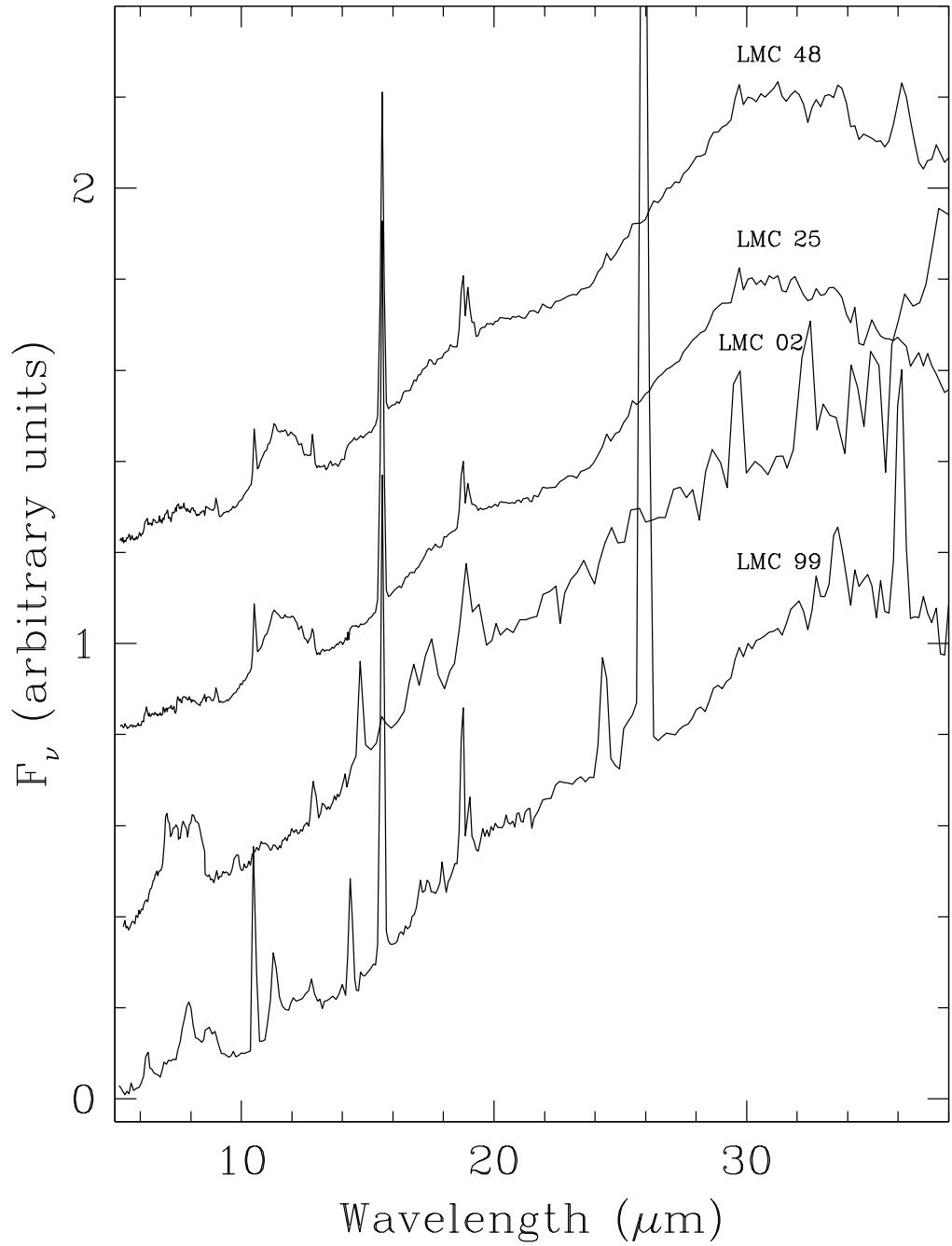}
\caption{Same as Figure 2 but for the four
Large Magellanic Cloud PNe in our sample. \label{fig3}}
\end{figure}

\clearpage

\begin{figure}
\includegraphics[angle=0,scale=.70]{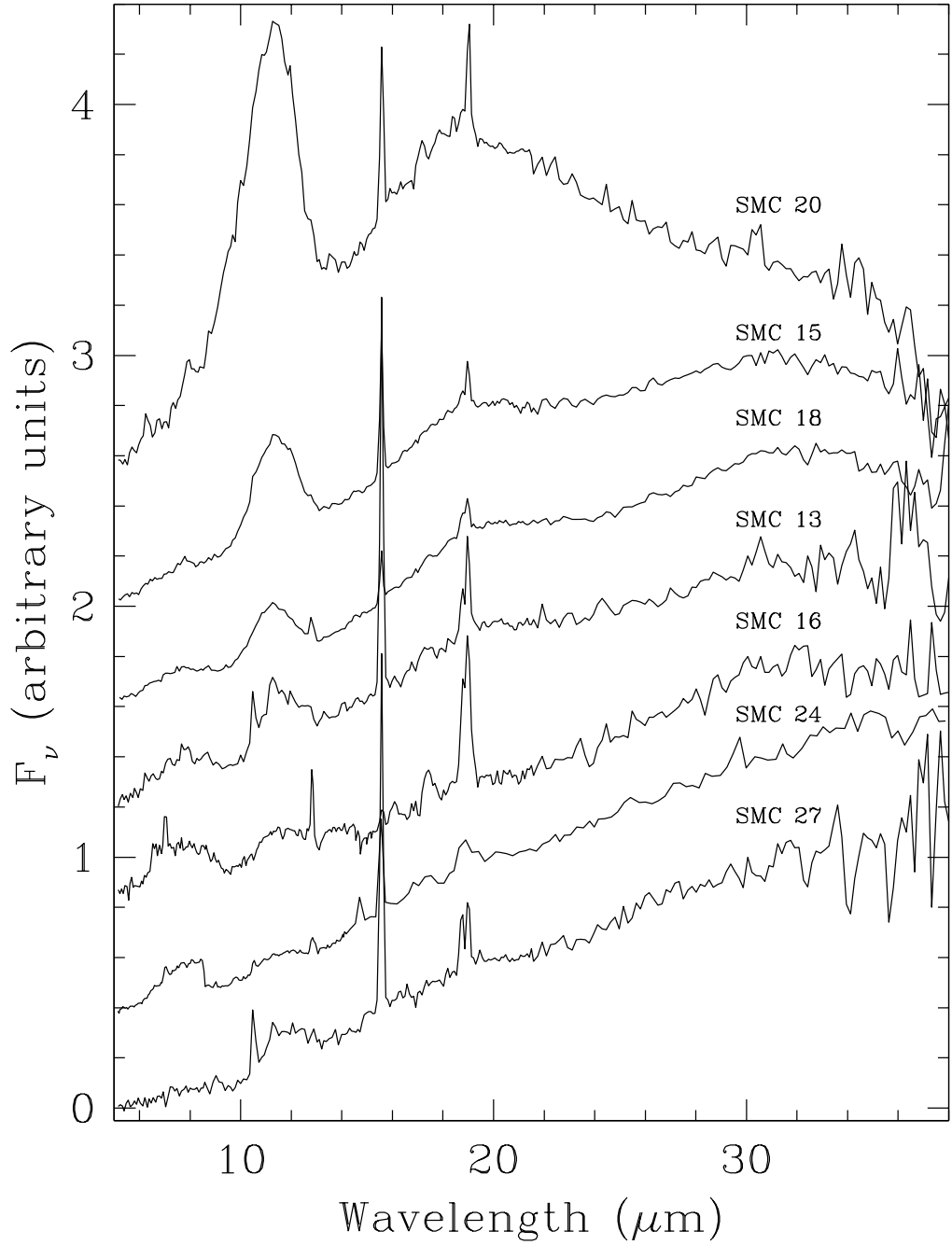}
\caption{Same as Figure 2 but for the seven
Small Magellanic Cloud PNe in our sample. \label{fig4}}
\end{figure}

\clearpage

\begin{figure}
\includegraphics[angle=0,scale=.70]{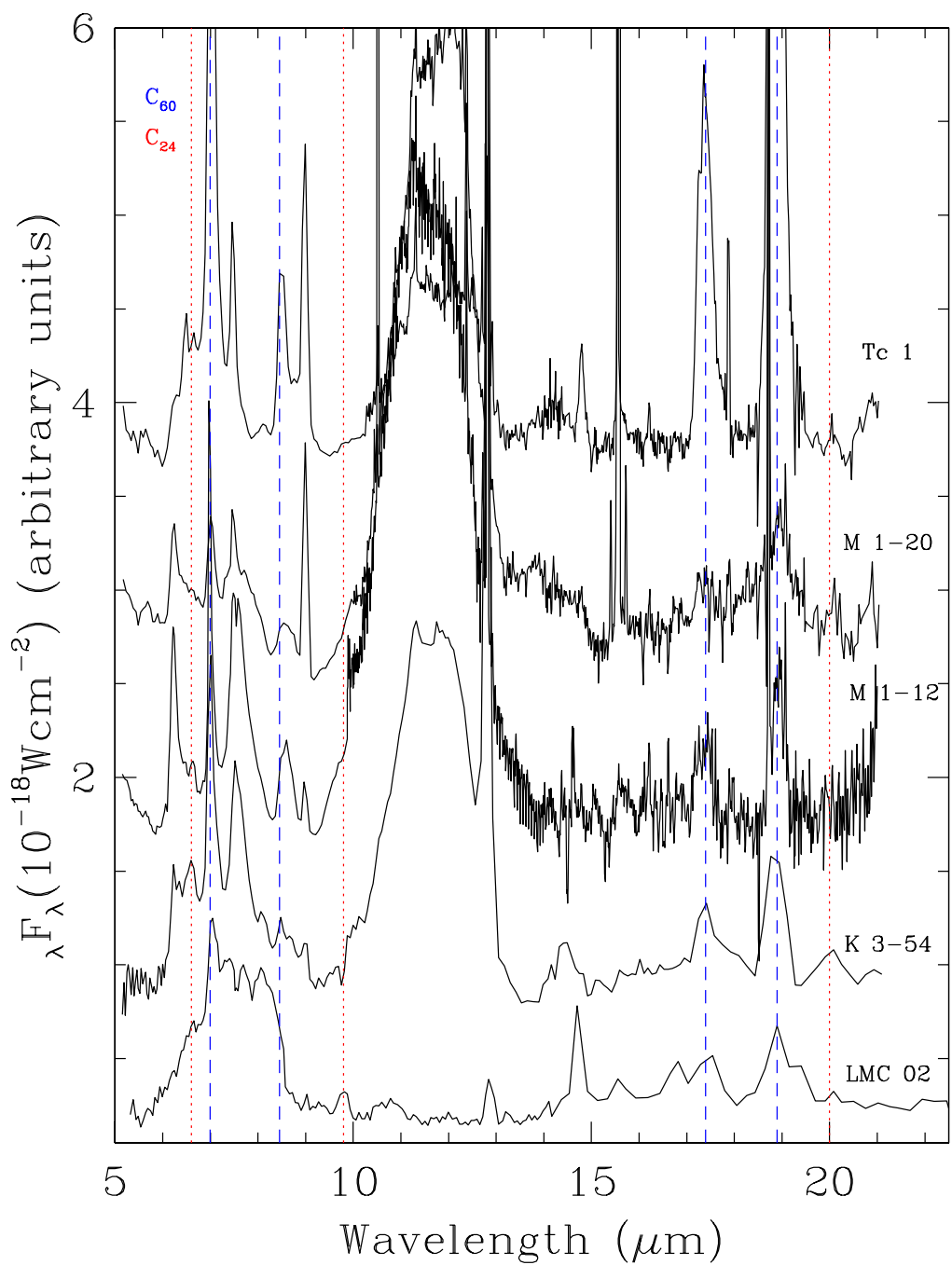}
\caption{Residual spectra in the wavelength range $\sim$5$-$23 $\mu$m for 
four Galactic fullerene PNe (Tc 1, M 1-20, M 1-12, and K 3-54) in
comparison with the extragalactic PN LMC 02. The band positions of C$_{60}$
(dashed) and planar C$_{24}$ (dotted) are marked (see text for more details).
Note that the spectra are artificially displaced in the vertical axis for clarity. \label{fig5}}
\end{figure}

\clearpage

\begin{figure}
\includegraphics[angle=90,scale=.70]{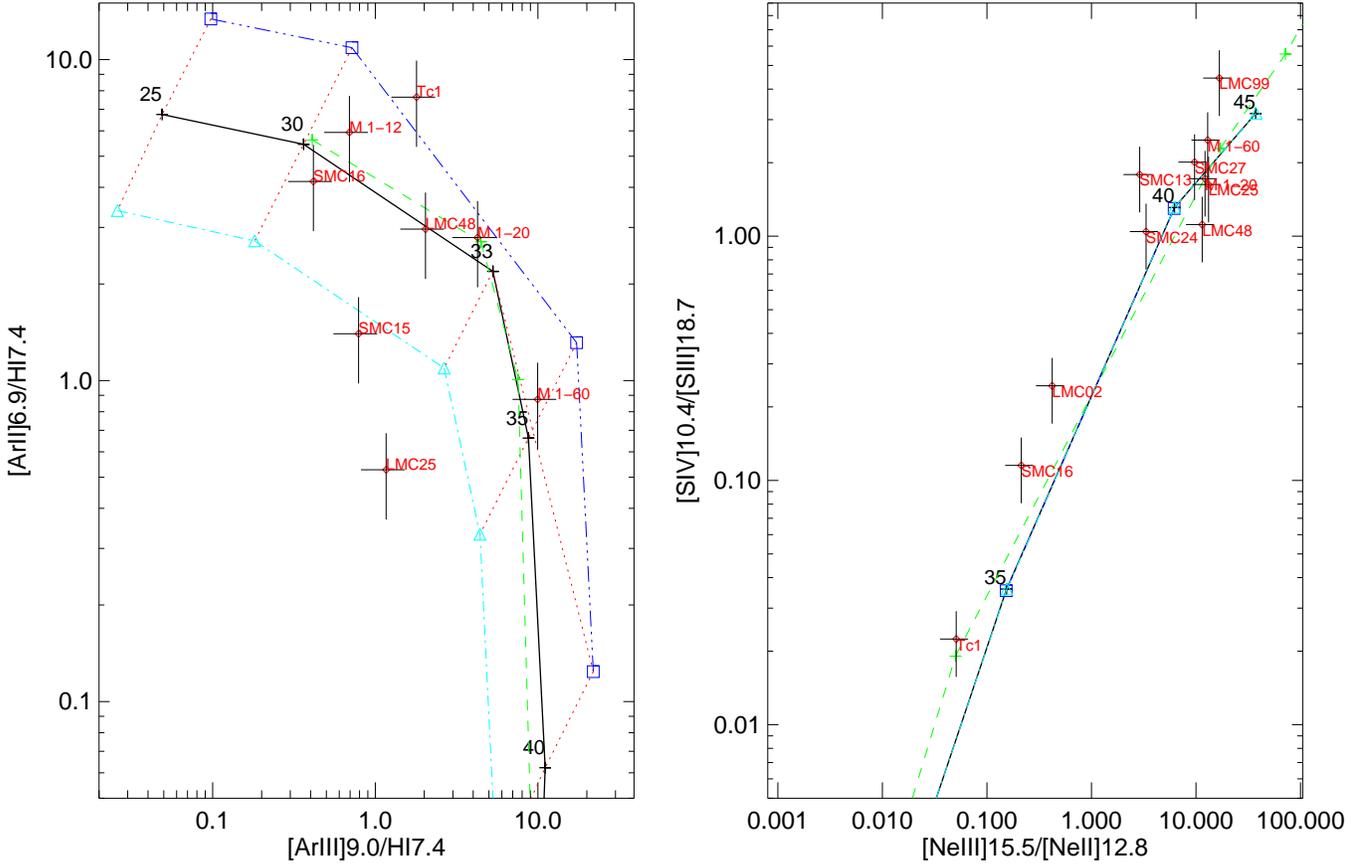}
\caption{Comparison of {\it CLOUDY} photoionization models with observed line ratios in
fullerene PNe. The black continuous line represents the WMBasic model described
in the text, using solar metallicity for the star atmosphere. The model
represented with the green line correspond to a sub-solar metallicity. The
number close to each point indicates the value of central star temperature in
units of 10$^{3}$ K. The models in the left panel displayed by dotted-dashed
(cyan) and dotted-dotted-dashed (blue) lines represent models with Ar abundance
a factor 2 below and above that one quoted by Pottasch et al. (2011),
respectively. The dotted lines join models with the same value of the star
temperature. Error bars of 40\% in the observed line intensity ratios are
also shown.
\label{fig6}}
\end{figure}

\clearpage

\begin{figure}
\includegraphics[angle=-90,scale=.35]{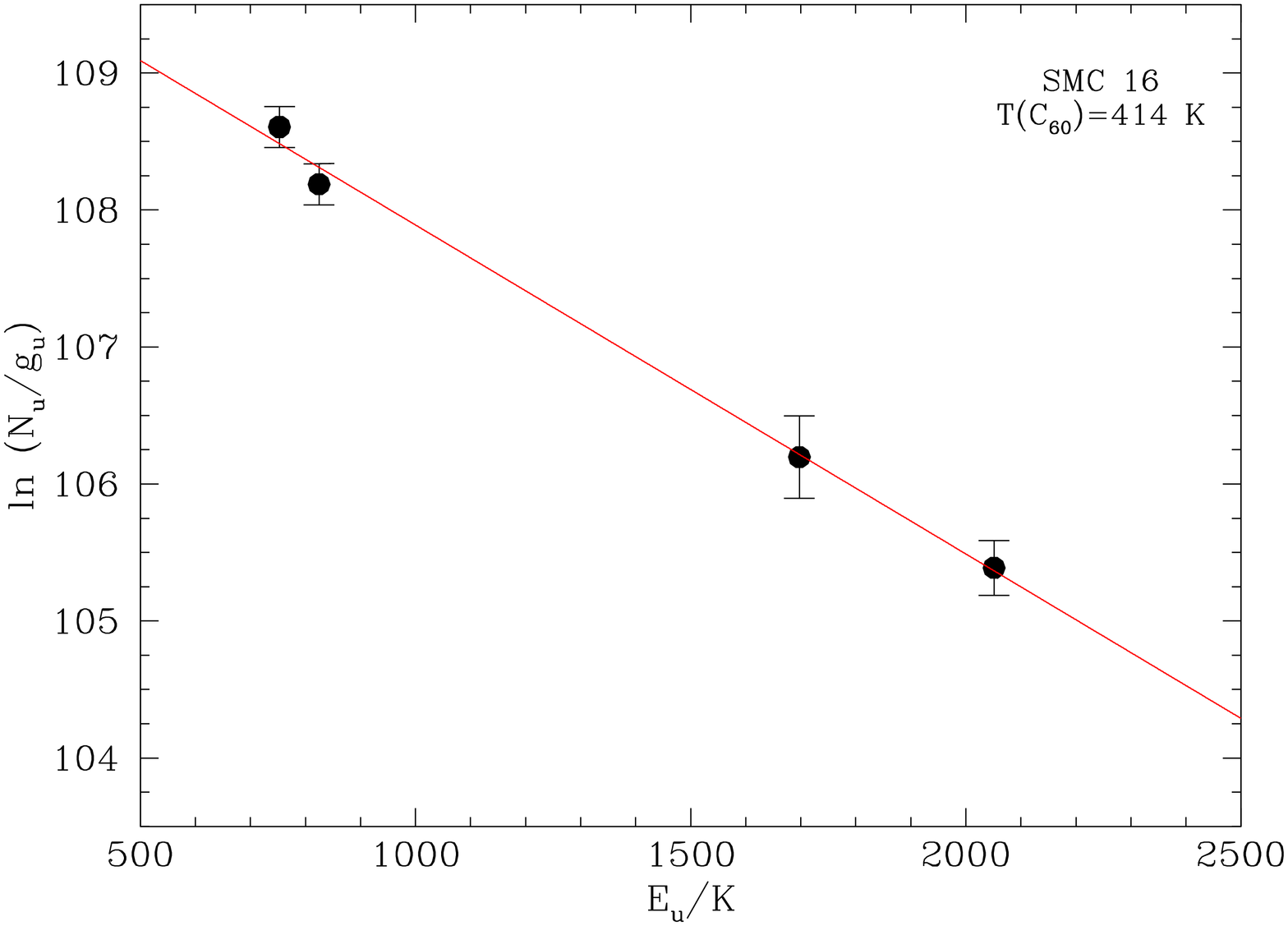}%
\includegraphics[angle=-90,scale=.35]{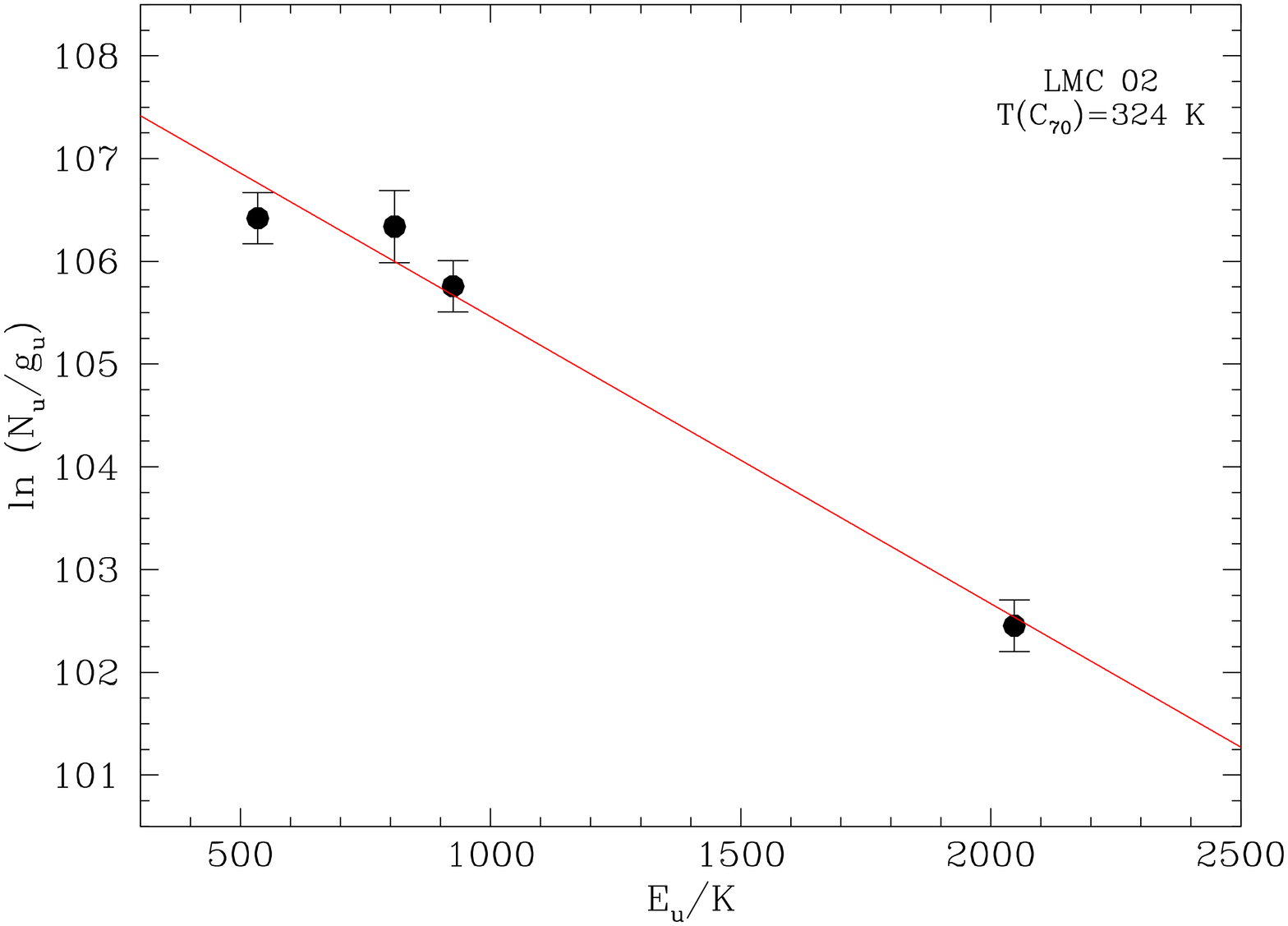}
\caption{Illustrative examples of the Boltzmann excitation diagrams
(ln(N$_{nu}$/g$_{u}$) vs. E$_{u}$/k) obtained for the C$_{60}$ and C$_{70}$
bands observed in the PNe SMC 16 (left panel) and LMC 02 (right
panel), respectively (see text). \label{fig7}}
\end{figure}

\clearpage

\begin{figure}
\includegraphics[angle=0,scale=.60]{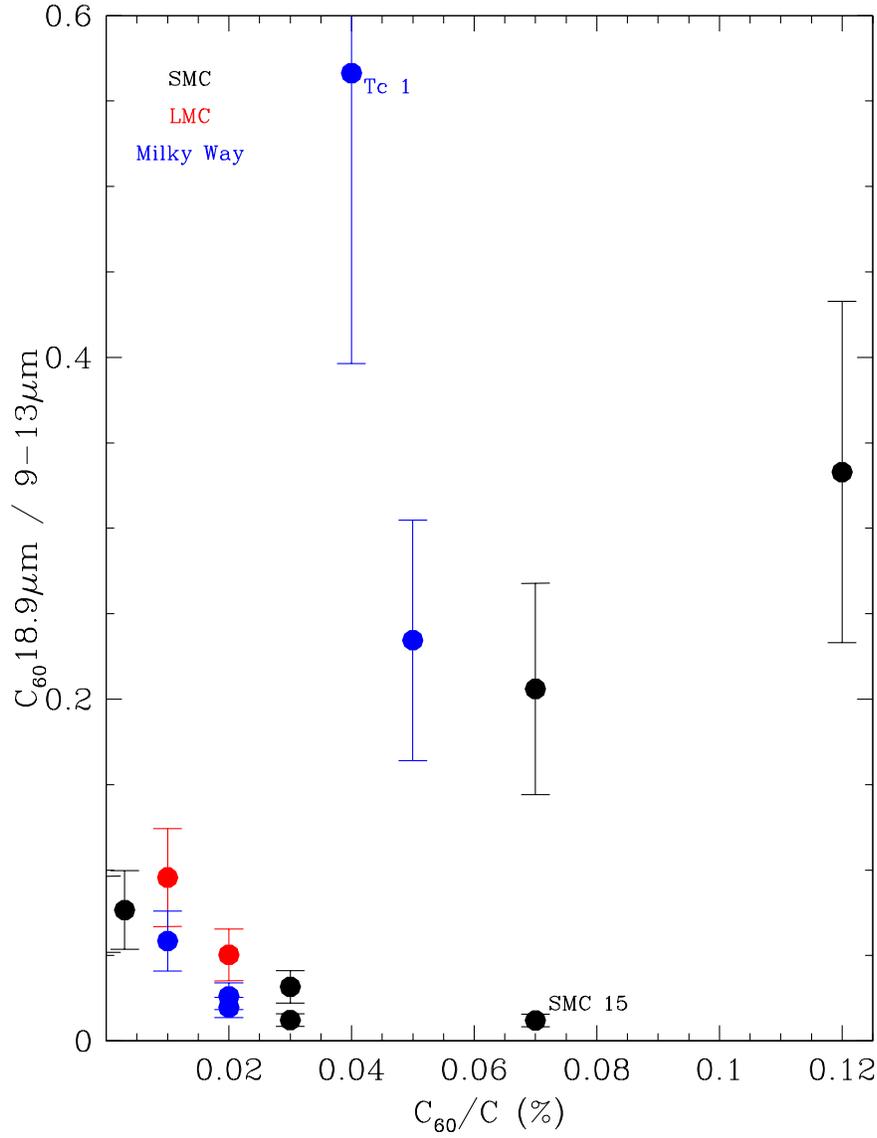}
\caption{C$_{60}$ 18.9$\mu$m/9$-$13$\mu$m flux ratio versus the C$_{60}$/C
abundance (in percentage) for fullerene PNe pertaining to different
environments such as the Galaxy and the Magellanic Clouds. Note that the
apparent two outliers Tc 1 and SMC 15 are indicated (see text for more details).}
\label{fig8}
\end{figure}

\clearpage

\begin{figure}
\includegraphics[angle=0,scale=.60]{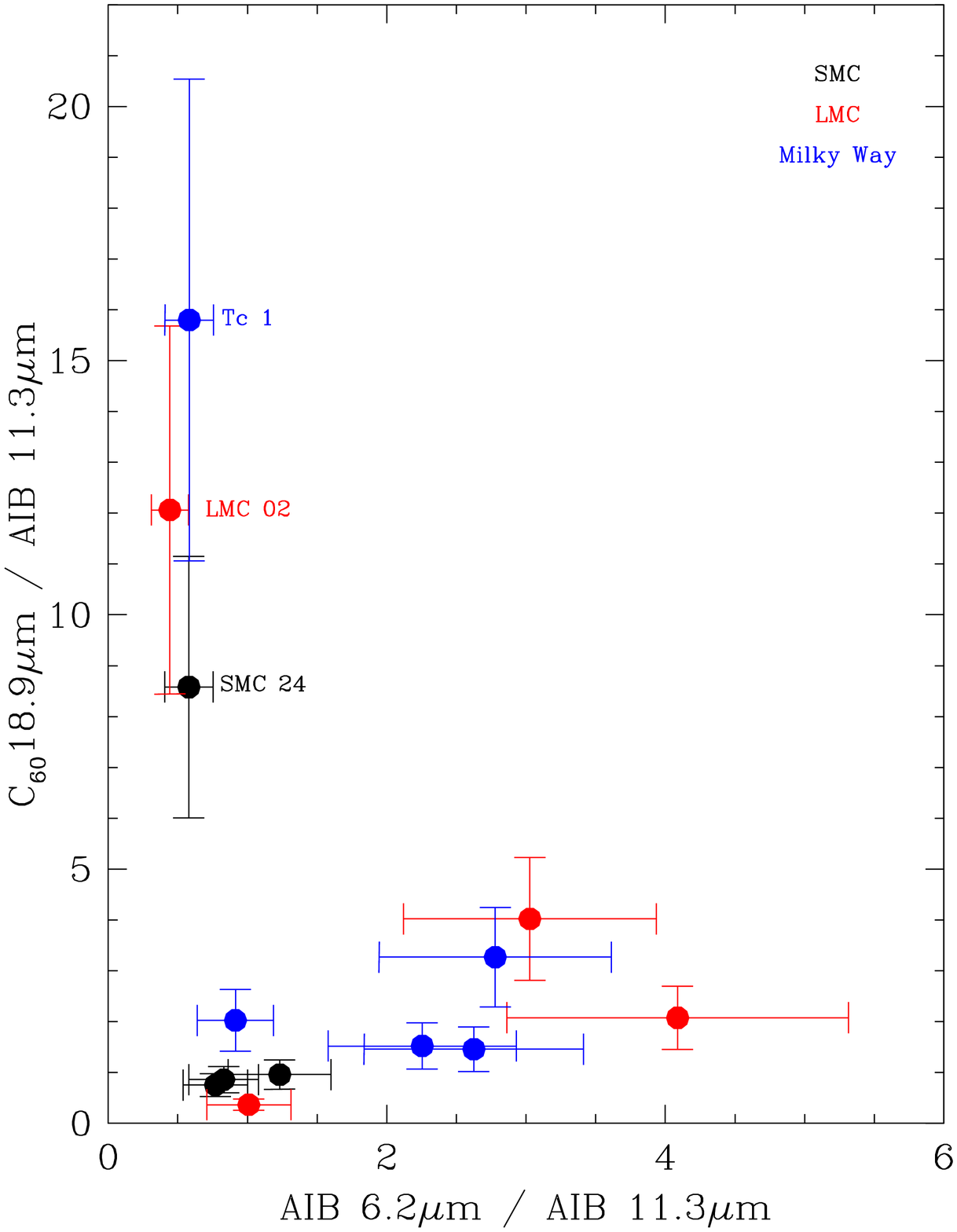}
\caption{C$_{60}$ 18.9$\mu$m/AIB 11.3$\mu$m flux ratio versus the AIB
6.2$\mu$m/AIB 11.3$\mu$m flux ratio for fullerene PNe pertaining to
different environments such as  the Galaxy and the Magellanic Clouds. Note
that several PNe are indicated.}
\label{fig9}
\end{figure}

\clearpage

\begin{figure}
\includegraphics[angle=0,scale=.60]{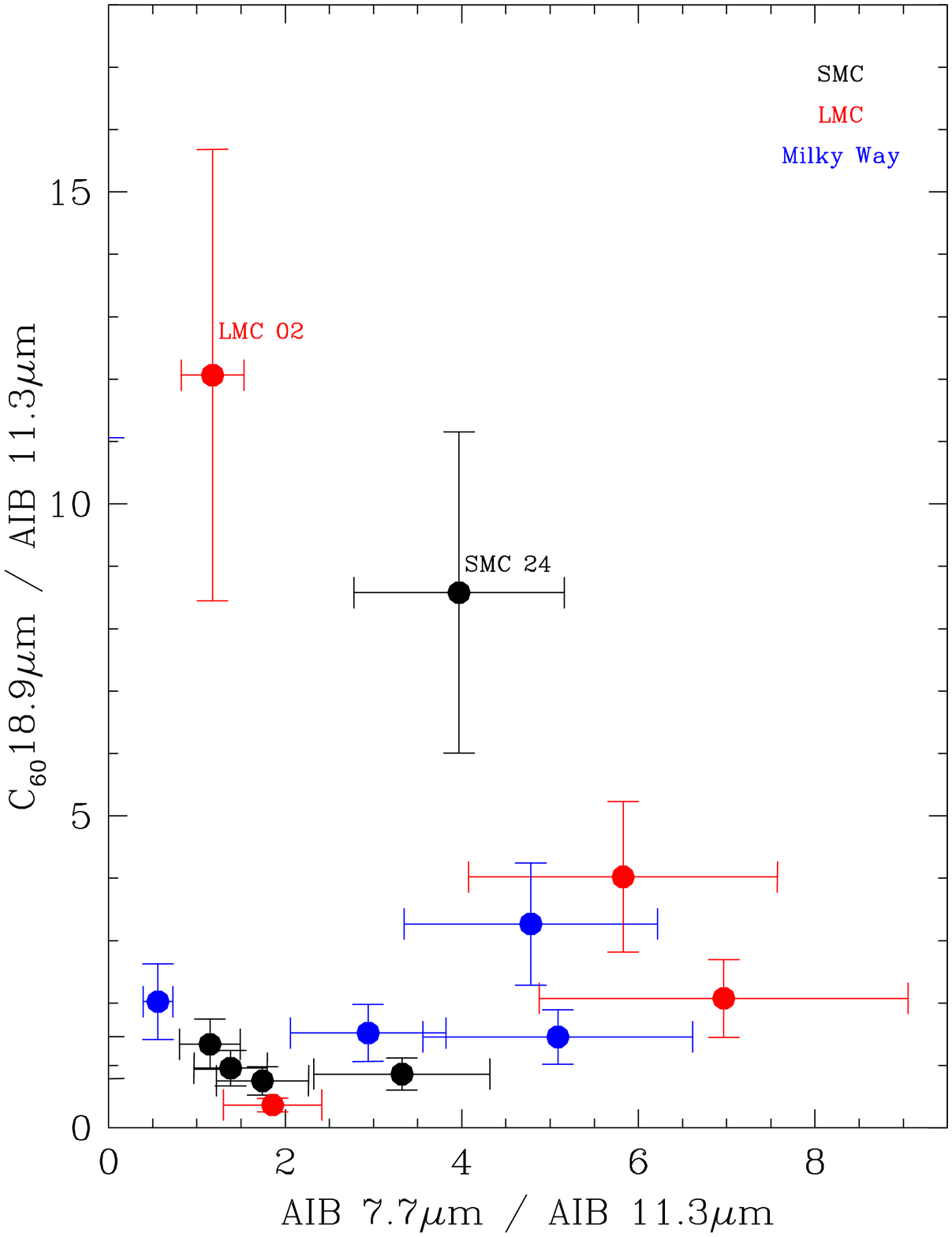}
\caption{C$_{60}$ 18.9$\mu$m/AIB 11.3$\mu$m flux ratio versus the AIB
7.7$\mu$m/AIB 11.3$\mu$m flux ratio for fullerene PNe pertaining to
different environments such as our own Galaxy and the Magellanic Clouds. Note
that several PNe are indicated.}
\label{fig10}
\end{figure}

\clearpage

\begin{figure}
\includegraphics[angle=0,scale=.60]{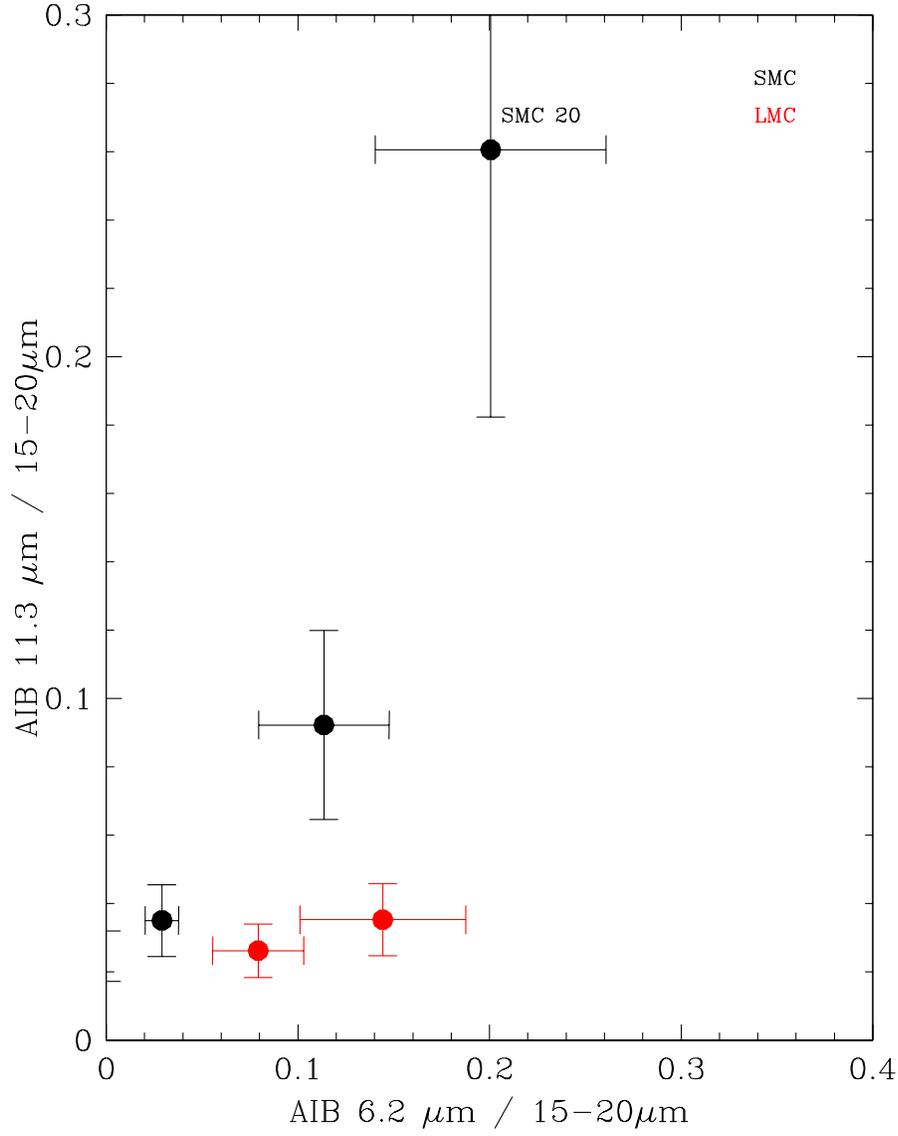}
\caption{AIB 11.3$\mu$m/15$-$20$\mu$m flux ratio versus the AIB 6.2$\mu$m/15$-$20$\mu$m 
flux ratio for fullerene PNe pertaining to the Magellanic Clouds.
Note that PN SMC 20 is indicated.}
\label{fig11}
\end{figure}

\end{document}